\documentclass{article}

\usepackage{authblk}
\usepackage[square]{natbib}
\usepackage{yhmath}
\usepackage{mathtools, amsmath, amsthm, amssymb, amsbsy, amsfonts, amscd}
\usepackage{euscript}
\usepackage{bm}
\usepackage{color, soul}
\usepackage{empheq}
\usepackage{rotating}
\usepackage{enumerate}
\usepackage{enumitem}
\usepackage{bbold}

\usepackage{ftnxtra}
\usepackage{fnpos}
\usepackage{appendix}

\usepackage{graphicx}
\usepackage{pstool}
\usepackage{psfrag}

\usepackage{comment}

\numberwithin{equation}{section}

\theoremstyle{plain}

\theoremstyle{definition}	
 
 \newtheorem{remark}{Remark}[section]

\usepackage{caption2}

\usepackage{hyperref}
\hypersetup{colorlinks=true, linkcolor=blue}
\hypersetup{colorlinks=true,citecolor=blue}

\DeclareMathAlphabet{\mathpzc}{OT1}{pzc}{m}{it}

\usepackage{mathrsfs}

\definecolor{lighter_purple_mathematica}{rgb}{0.6666666666,0.33333333333,0.666666666666}

\makeatletter
\newsavebox{\@brx}
\newcommand{\llangle}[1][]{\savebox{\@brx}{\(\m@th{#1\langle}\)}%
  \mathopen{\copy\@brx\mkern2mu\kern-0.9\wd\@brx\usebox{\@brx}}}
\newcommand{\rrangle}[1][]{\savebox{\@brx}{\(\m@th{#1\rangle}\)}%
  \mathclose{\copy\@brx\mkern2mu\kern-0.9\wd\@brx\usebox{\@brx}}}%
\let\oldabs\abs
\def\abs{\@ifstar{\oldabs}{\oldabs*}}
\makeatother

\usepackage{accents}
\newcommand{\Fe}{\accentset{e}{\mathbf{F}}}

\newcommand{\Fg}{\accentset{g}{\mathbf{F}}}

\newcommand{\cFg}{\accentset{g}{F}}

\newcommand{\Jg}{\accentset{g}{J}}
\newcommand{\Je}{\accentset{e}{J}}

\newcommand{\We}{\accentset{e}{W}}
\newcommand{\Wg}{\accentset{g}{W}}

\usepackage[all,cmtip]{xy}
\usepackage{xypic}

\begin{document}

\title{\textbf{Nonlinear Mechanics of Arterial Growth}}

\author[1]{Aditya Kumar}
\author[1,2]{Arash Yavari\thanks{Corresponding author, e-mail: arash.yavari@ce.gatech.edu}}
\affil[1]{\small \textit{School of Civil and Environmental Engineering, Georgia Institute of Technology, Atlanta, GA 30332, USA}}
\affil[2]{\small \textit{The George W. Woodruff School of Mechanical Engineering, Georgia Institute of Technology, Atlanta, GA 30332, USA}}

\maketitle

\begin{abstract}
In this paper, we formulate a geometric theory of the mechanics of arterial growth. An artery is modeled as a finite-length thick shell that is made of an incompressible nonlinear anisotropic solid. An initial radially-symmetric distribution of finite radial and circumferential eigenstrains is assumed. Bulk growth is assumed to be isotropic. A novel framework is proposed to describe the time evolution of growth, governed by a competition between the elastic energy and a \emph{growth energy}. The governing equations are derived through a two-potential approach and using the Lagrange-d'Alembert principle. An isotropic dissipation potential is considered, which is assumed to be convex in the rate of growth function. Several numerical examples are presented that demonstrate the effectiveness of the proposed model in predicting the evolution of arterial growth and the intricate interplay among eigenstrains, residual stresses, elastic energy, growth energy, and dissipation potential.
A distinctive feature of the model is that the growth variable is not constrained by an explicit upper bound; instead, growth naturally approaches a steady-state value as a consequence of the intrinsic energetic competition.
\end{abstract}

\begin{description}
\item[Keywords:] Nonlinear elasticity, material growth, mechanics of bulk growth, anelasticity.
\end{description}

\tableofcontents

\section{Introduction}

The growth of arteries in response to changes in lumen pressure, which cause the vessel walls to thicken, is of significant interest. Such a response falls under the general category of mechanically influenced growth. Residual stresses and strains are generated in arterial walls during growth. It has been known since the 1800s \citep{Wolff1870, woods1892heart, thoma1893artery} that excess mechanical stimuli trigger growth in biological tissues.  However, the fundamental physics of this process remains poorly understood at the tissue level.

Mechanics of bulk growth has been a subject of increasing interest in the past few decades. The first studies of bulk growth inspired by applications in the field of biology appeared in the 1980s and 1990s \citep{Skalak1982,Fung1983,KondaurovNikitin1987,TakamizawaMatsuda1990,Takamizawa1991,Rodriguez1994}.
There exist numerous theoretical and computational studies on the mechanics of bulk growth; see \citep{TakamizawaMatsuda1990,Takamizawa1991,Rodriguez1994,EpsteinMaugin2000,LubardaHoger2002,Garikipati2004,BenAmarGoriely2005,Himpel2005,Klarbring2007,Yavari2010,Goktepe2010, kuhl2012skin, kuhl2014artery, Sadik2016} and references therein.
Traditionally, in the kinematics of bulk growth, in addition to the standard deformation map, an independent field(s) describing bulk growth is introduced. A challenging ingredient of any growth theory is the kinetic equation(s) governing the dynamics of bulk growth.
Kinetic equation of bulk growth (or ``evolution laws" \citep{Goriely2017}) are usually postulated based on certain symmetry assumptions. Examples are: isotropic growth, orthotropic growth, transversely isotropic growth, etc. \citep{Goriely2017}. Starting from a multiplicative decomposition of deformation gradient into elastic and growth distortions $\mathbf{F}=\Fe\Fg$,\footnote{For a detailed history of this decomposition see \citep{Sadik2017,YavariSozio2023}.} a growth equation has the following form: $\dot{\Fg}=\boldsymbol{\mathsf{G}}(X,\mathbf{C}^{\flat},\mathbf{G})$. Note that $\Fg:T_X\mathcal{B}\to T_X\mathcal{B}$ is a material tensor, and hence is automatically objective. It should also be noted that the material metric explicitly depends on $\Fg$: $\mathbf{G}=\Fg^*\mathring{\mathbf{G}}=\Fg^\star\mathring{\mathbf{G}}\Fg$, where $\mathring{\mathbf{G}}$ is the flat metric of the Euclidean space, or more precisely, the metric of the stress-free body (in the absence of growth) induced from the Euclidean ambient space.
The precise functional form of $\mathbf{G}$ has remained unclear and has been the focus of extensive investigation.

It has been known for quite some time that arteries are residually stressed \citep{Fung2013,Vaishnav1987}.
\citet{Takamizawa1991} proposed that the stress-free configuration of a residually-stressed artery can be modeled by a Riemannian manifold. They pointed out that the Riemannian metric is closely related to \emph{residual strain} (or what is usually called eigenstrain\footnote{The hybrid German–English term \emph{eigenstrain} originates in the pioneering work of Hans Reissner \citep{Reissner1931}, where \emph{Eigenspannung} denotes a proper or self strain, and was subsequently popularized by Mura \citep{Kinoshita1971,Mura1982}. In the literature, several equivalent notions appear under different names, including \emph{initial strain} \citep{Kondo1949}, \emph{nuclei of strain} \citep{Mindlin1950}, \emph{transformation strain} \citep{Eshelby1957}, \emph{inherent strain} \citep{Ueda1975}, and \emph{residual strains} \citep{ambrosi2019growth} (see also \citep{Jun2010,Zhou2013}).} in the literature).
\citet{Takamizawa1987} assumed a uniform circumferential strain for arteries under physiological loading conditions. They showed that this assumption leads to almost uniform stress distributions and non-vanishing residual stresses.
\citet{Rachev1999} modeled an artery as a thick-walled tube made of an incompressible orthotropic  elastic solid \citep{Patel1969}. In addition to elastic (passive) stresses determined constitutively (up to an unknown pressure field), they assumed an active circumferential stress due to muscle contraction and relaxation. In their numerical examples, they observed that eigenstrains (referred to as ``residual strains") are highly sensitive to muscle contraction and relaxation.
There have been several other studies of the mechanics of arterial growth and remodeling in the literature \citep{Holzapfel2002,Rodriguez2007,Cardamone2009}.

\citet{Saez2014} presented a computational framework for modeling hypertensive growth in the human carotid artery, focusing on the thickening of the arterial wall due to smooth muscle cell hypertrophy. Using a finite element implementation of finite growth based on a multiplicative decomposition of the deformation gradient, they investigated how mechanical stimuli, particularly stretch, influence growth over time. This study involves several important modeling assumptions, including the choice of growth kinematics, the definition of mechanical homoestasis (based on stress or strain), and the layer-specific growth behavior of arterial tissues. The model is applied to both idealized cylindrical geometries and patient-specific carotid artery reconstructions, showing good agreement with experimental observations.


Recently, \cite{ErlichZurlo2025} presented a geometric reformulation of biological growth in which the evolution of growth is driven not only by the conventional homeostatic-stress mechanism but also by deviations of the material manifold’s Ricci curvature from a prescribed target. They proposed that growth and remodeling in living tissues aim to reach a homeostatic state characterized by a physiological level of geometric frustration quantified through the Ricci curvature of the material manifold. In this framework, incompatibility---measured by the Ricci tensor---emerges as the fundamental geometric quantity linking curvature, growth, and residual stress. To formalize this idea, they introduced a ``growth action'' functional analogous to the Einstein--Hilbert action, which penalizes deviations from a target curvature and yields a corresponding expression for the homeostatic Eshelby stress tensor. The growth equation governing the rate of change of the growth tensor is then postulated to depend on the difference between the actual and target Eshelby stresses, rather than being derived from thermodynamic principles.

It remains unclear what it means for a living system to have mechanical homeostasis. As discussed above,  the dominant view of mechanical homeostasis is that the body grows and remodels to maintain a homeostatic value of stress or strain \citep{Lubarda2002, kuhl2014review}. This view has been developed by drawing inspiration from homeostatic physiological values of temperature and pH in the body. While this could be a reasonable approach for biological structures in which a nearly uniform and largely uniaxial state of stress and strain exists, for instance, in the muscle tissue, in general, biological structures have a triaxial non-uniform stress state. Arteries are an example of such a structure. Under such non-uniform stress and strain states, the very notion of mechanical homeostasis becomes ambiguous.

Our motivation in this paper is to develop a systematic and self-consistent framework for modeling the evolution of growth in living tissues. A central question is whether experimentally observed phenomena, such as changes in arterial wall diameter and thickness under altered physiological conditions, can be quantitatively described within a rigorous mechanical and mathematical theory of growth and remodeling. In this paper, we formulate the mechanics of arterial growth within a two-potential framework based on the Lagrange--d'Alembert principle, which naturally incorporates both energetic and dissipative effects associated with growth-induced deformation.
A key feature of our model is that the growth variable is not restricted by an explicit upper bound but instead evolves toward a steady-state value as a result of the intrinsic energetic competition.

This paper is organized as follows. In \S\ref{Sec:Kinematics}, we develop a general variational formulation for bulk growth in anisotropic solids based on the Lagrange--d'Alembert principle and introduce a notion of growth energy which, together with the elastic energy and the dissipation potential, governs the dynamics of bulk growth. The theory is then specialized in \S\ref{Sec:Arteries} to model isotropic arterial growth, where we derive the governing equations and examine their structure. In \S\ref{Sec:NumericalResults}, we present numerical results that illustrate the coupled evolution of deformation, growth, and residual stress under both physiological and pathological loading conditions. Finally, in \S\ref{Sec:Conclusions}, we summarize the main results and discuss possible extensions of the framework.

\section{Bulk Growth} \label{Sec:Kinematics}

In this section we formulate a general geometric variational theory of bulk growth. The framework will then be specialized to the case of arterial growth in \S\ref{Sec:Arteries}.

\subsection{Kinematics}

\paragraph{Motion, reference, and current configurations.}
Let us consider a growing body $\mathcal{B}$ that while moves in the Euclidean ambient space $\mathcal{S}$ undergoes bulk growth. The Euclidean metric in the ambient space is denoted by $\mathbf{g}$.
The body is an embedded $3$-submanifold of the Euclidean ambient space $\mathcal{S}$. The initial embedding of the body into the Euclidean ambient space induces a flat metric $\mathring{\mathbf{G}}=\mathbf{g}\big|_{\mathcal{B}}$ on the body. This is the natural metric of the elastic body before the body experiences any growth.
Motion of the growing body is a one-parameter family of mappings $\varphi_t:\mathcal{B}\to\mathcal{C}_t\subset\mathcal{S}$, where $\mathcal{C}_t=\varphi_t(\mathcal{B})$ is the current configuration of the growing body (more precisely, motion is a curve $t\mapsto\varphi_t$ in the space of all configurations of $\mathcal{B}$). A material point $X\in\mathcal{B}$ is mapped to $x_t=x(X,t)=\varphi_t(X)\in\mathcal{C}_t$. 
The Riemannian manifolds $(\mathcal{B},\mathring{\mathbf{G}})$ and $(\mathcal{S},\mathbf{g})$ are the initial material manifold and the ambient space manifold, respectively.

The derivative of the deformation map is usually called deformation gradient and is denoted by $\mathbf{F}(X,t)=T\varphi_t(X):T_X\mathcal{B}\to T_x\mathcal{C}_t$, where $T_X\mathcal{B}$ and $T_x\mathcal{C}_t$ are the tangent spaces of $\mathcal{B}$ at $X$ and $\mathcal{C}_t$ at $x_t$, respectively. 
Let us consider the local coordinate charts $\{X^A\}$ and $\{x^a\}$ for $\mathcal{B}$ and $\mathcal{C}$, respectively. $\mathbf{F}$ has the following coordinate representation 
\begin{equation} 
	\mathbf{F}(X,t)=\frac{\partial\varphi^a(X,t)}{\partial X^A}\,\frac{\partial}{\partial x^a}
	\otimes dX^A
	\,.
\end{equation}
Its adjoint $\mathbf{F}^{\star}(X,t): T^*_x\mathcal{C}_t\to T^*_X\mathcal{B}$, where $T^*_X\mathcal{B}$ and $T^*_x\mathcal{C}_t$ are the co-tangent spaces of $\mathcal{B}$ and $X$ and $\mathcal{C}_t$ at $x$, respectively, and $\langle .,. \rangle$ is the natural paring of $1$-forms and vectors, has the following coordinate representation 
\begin{equation} 
	\mathbf{F}^{\star}(X,t)=\frac{\partial\varphi^a(X,t)}{\partial X^A} \,dX^A 
	\otimes \frac{\partial}{\partial x^a}\,.
\end{equation}
The transpose of the deformation gradient $\mathbf{F}^{\mathsf{T}}(X,t): T_x\mathcal{C}_t\to T_X\mathcal{B}$ is metric dependent and has components $\big(F^{\mathsf{T}}\big)^A{}_a=G^{AB}F^b{}_B\,g_{ba}$, or $\mathbf{F}^{\mathsf{T}}=\mathbf{G}^{\sharp}\mathbf{F}^{\star}\mathbf{g}$. 
The right Cauchy-Green strain is defined as $\mathbf{C}^{\flat}=\varphi^*\mathbf{g}=\mathbf{F}^*\mathbf{g}=\mathbf{F}^{\star}\mathbf{g}\mathbf{F}$. In components, $C_{AB}=F^a{}_A\,g_{ab}F^b{}_B$. Also, notice that
\begin{equation} 
	C^A{}_B=G^{AM}C_{MB}=(G^{AM}F^a{}_M\,g_{ab})F^b{}_B
	=\big(F^{\mathsf{T}}\big)^A{}_b\,F^b{}_B
	\,,
\end{equation}
i.e., $\mathbf{C}=\mathbf{F}^{\mathsf{T}}\mathbf{F}$.
The spatial analogue of the right Cauchy-Green strain is defined as $\mathbf{c}^{\flat}=\mathbf{F}_*\mathbf{G}=\mathbf{F}^{-\star}\mathbf{G}\mathbf{F}^{-1}$. It has components, $c_{ab}=F^{-A}{}_a\,G_{AB}\,F^{-B}{}_b$, where $F^{-A}{}_a$ are components of $\mathbf{F}^{-1}$.
The left Cauchy-Green strain is defined as $\mathbf{B}^{\sharp}=\varphi^*\mathbf{g}^{\sharp}$, with components $B^{AB}=F^{-A}{}_a\,F^{-B}{}_b\,g^{ab}$. Its spatial analogue is defined as $\mathbf{b}^{\sharp}=\varphi_*\mathbf{G}^{\sharp}=\mathbf{F}\mathbf{G}^{\sharp}\mathbf{F}^{\star}$, which has  components, $b^{ab}=F^a{}_AF^b{}_B\,G^{AB}$. It is straightforward to show that $\mathbf{B} = \mathbf{C}^{-1}$ and $\mathbf{b} = \mathbf{c}^{-1}$.

We assume a multiplicative decomposition of the deformation gradient $\mathbf{F}(X,t)=\Fe(X,t)\,\Fg(X,t)$, where $\Fg(X,t):T_X\mathcal{B}\to T_X\mathcal{B}$ is a material tensor while $\Fe(X,t):T_X\mathcal{B}\to T_x\mathcal{C}$ is a two-point tensor. The natural distances in the growing body are measured using the material metric $\mathbf{G}=\Fg^*\mathring{\mathbf{G}}=\Fg^\star\mathring{\mathbf{G}}\Fg$.\footnote{See \citep{Sadik2017} and \citep{YavariSozio2023} for detailed discussions and literature review of the multiplicative decomposition in anelasticity.} In components, $G_{AB}=\cFg^M{}_A\,\mathring{G}_{MN}\,\cFg^N{}_B$.
The natural volume element of the Riemannian manifold $(\mathcal{B},\mathbf{G})$ at $X\in\mathcal{B}$ is denoted by $dV(X)$. The corresponding volume element in the current configuration at $x=\varphi(X)\in\mathcal{C}$ is denoted by $dv(x)$.
The Jacobian of the deformation relates the deformed and undeformed Riemannian volume elements as $dv(x)=JdV(X)$, where
\begin{equation}
	J=\sqrt{\frac{\det\mathbf{g}}{\det\mathbf{G}}}\det\mathbf{F}\,.
\end{equation}
Let us denote the Levi-Civita connections associated with the metrics $\mathbf{G}$ and $\mathbf{g}$ by $\nabla^{\mathbf{G}}$ and $\nabla^{\mathbf{g}}$, respectively.

\subsection{Constitutive equations of a growing anisotropic body}

In this section the constitutive equations of an isotropic growing solid are briefly reviewed. More specifically, the elastic energy function and dissipation potential are discussed. Next a growth energy is introduced.

\subsubsection{Energy function of an anisotropic growing body}

It is assumed that the growing body is made of a hyper-anelastic material, i.e., it has an elastic energy function density per unit mas that explicitly depends on the elastic distortion: $\We=\We(X,\Fe,\mathring{\boldsymbol{\Lambda}},\mathring{\mathbf{G}},\mathbf{g})$, where $\mathring{\boldsymbol{\Lambda}}$ is a set of structural tensors that explicitly depends on the symmetry group of the material \citep{liu1982,boehler1987,zheng1993,zheng1994theory,Lu2000}. 
Recall that $\Fe=\Fg_*\mathbf{F}$.
When the structural tensors are included as arguments of the energy function, the elastic energy becomes materially covariant. In particular, this implies that
\begin{equation} \label{Elastic-Energy}
	\We(X,\Fe,\mathring{\boldsymbol{\Lambda}},\mathring{\mathbf{G}},\mathbf{g})=\We(X,\Fg_*\mathbf{F},\mathring{\boldsymbol{\Lambda}},\mathring{\mathbf{G}},\mathbf{g}) 
	=\We(X,\Fg^*\Fg_*\mathbf{F},\Fg^*\mathring{\boldsymbol{\Lambda}},
	\Fg^*\mathring{\mathbf{G}},\mathbf{g})
	=\We(X,\mathbf{F},\boldsymbol{\Lambda},\mathbf{G},\mathbf{g})
	\,,
\end{equation}
where 
\begin{equation} \label{Material-Metric}
	\mathbf{G}=\Fg^*\mathring{\mathbf{G}}=\Fg^\star\mathring{\mathbf{G}}\Fg
	\,,
\end{equation}
is the material metric and $\boldsymbol{\Lambda}=\Fg^*\mathring{\boldsymbol{\Lambda}}$ \citep{YavariSozio2023}.
Objectivity implies that $W=\hat{W}(X,\mathbf{C}^{\flat},\boldsymbol{\Lambda},\mathbf{G})$, where $\mathbf{C}^{\flat}=\mathbf{F}^*\mathbf{g}=\mathbf{F}^{\star}\mathbf{g}\mathbf{F}$.
Thus, one concludes that the elastic energy function of an anisotropic growing body is identical to its initial elastic energy function if one replaces the flat initial material metric $\mathring{\mathbf{G}}$ by the (evolving) material metric $\mathbf{G}$ and the structural tensors $\mathring{\boldsymbol{\Lambda}}$ by $\boldsymbol{\Lambda}$ \citep{YavariSozio2023}.

\paragraph{Measures of stress.}
In nonlinear anelasticity, several distinct tensorial measures of stress are used to describe the same physical quantity. Among the infinitely many admissible definitions, three are particularly useful because of their geometric clarity and their convenience in constitutive modeling. Consider an oriented surface element with area $da$ in the deformed configuration $\mathcal{C}$ and $\mathbf{g}$-unit normal $\mathbf{n}$. 
The traction acting on this element is given by $\mathbf{t}=\boldsymbol{\sigma}\mathbf{n}^\flat$, where $\boldsymbol{\sigma}$ denotes the \textit{Cauchy stress tensor}, $\mathbf{n}^\flat=\mathbf{g}\mathbf{n}$ (the $1$-form corresponding to $\mathbf{n}$) and the corresponding elemental force is $\mathbf{f}=\mathbf{t}\,da$. In components, this gives $t^a=\sigma^{ab}n_b$, with $n_b=g_{bc}n^c$. The corresponding area element in the reference configuration $\mathcal{B}$ has area $dA$ and $\mathbf{G}$-unit normal $\mathbf{N}$. The \textit{first Piola--Kirchhoff stress tensor} $\mathbf{P}$ is defined by relating forces in the deformed and reference configurations as
\begin{equation}
	\mathbf{t}\,da = \mathbf{P}\,\mathbf{N}^\flat\,dA\,,\qquad \mathbf{N}^\flat=\mathbf{G}\mathbf{N}
    \,.
\end{equation}
Using Nanson’s formula $\mathbf{n}^{\flat}\,da=J\,\mathbf{F}^{-\star}\,\mathbf{N}^{\flat}\,dA$ ($n_ada=J\,F^{-A}{}_a\,N_AdA$), one obtains the classical relation
\begin{equation}
	\mathbf{P} = J\,\boldsymbol{\sigma}\,\mathbf{F}^{-\star}\,,
\end{equation}
where the Jacobian is given by\footnote{For isotropic growth, one has $\sqrt{I_3} = J = \mathfrak{g}^{-3}\,\sqrt{\frac{\det\mathbf{g}}{\det\mathring{\mathbf{G}}}}\,\det\mathbf{F}
= \mathfrak{g}^{-3}\,\mathring{J}$, where $\mathring{J}$ denotes the Jacobian in the absence of growth.}
\begin{equation} \label{Jacobian-Elasticity}
	J=\sqrt{\frac{\det\mathbf{g}}{\det\mathbf{G}}}\,\det\mathbf{F}\,.
\end{equation}
In component form, $P^{aA}=J\,\sigma^{ab}\,F^{-A}{}_b$. Pulling the force $\mathbf{f}$ back to the reference configuration defines the \textit{second Piola--Kirchhoff stress tensor} $\mathbf{S}$ through $\mathbf{F}^{-1}\mathbf{t}\,da=\mathbf{S}\,\mathbf{N}^\flat\,dA$. It then follows that
\begin{equation}
	\mathbf{S} = \mathbf{F}^{-1}\mathbf{P} = J\,\mathbf{F}^{-1}\,\boldsymbol{\sigma}\,\mathbf{F}^{-\star}\,,
\end{equation}
or, equivalently in components,
$S^{AB}=F^{-A}{}_a\,P^{aB}=J\,F^{-A}{}_a\,\sigma^{ab}\,F^{-B}{}_b$.
For a hyper-anelastic body, the first Piola–Kirchhoff stress tensor is obtained from the deformation through the energy function as
\begin{equation}
	\mathbf{P} = \mathbf{g}^\sharp\frac{\partial \We}{\partial \mathbf{F}} 
    \,.
\end{equation}
Equivalently,
\begin{equation}
	\boldsymbol{\sigma}
    =J^{-1}\mathbf{g}^\sharp\frac{\partial \We}{\partial \mathbf{F}}\mathbf{F}^\star\,.
\end{equation}

\begin{remark}
In the classical approach to growth mechanics, elastic energy is defined with respect to the ``intermediate configuration". The volume element of the intermediate configuration $d\mathring{V}$ and that of the material manifold $dV$ are related as $d\mathring{V}=\Jg dV$, where $\Jg=\det\Fg$. Let us denote the classical elastic energy by $\We_0$. Thus, $\We_0\,d\mathring{V}=\We dV$, and hence, $\We_0=\Jg^{-1} \We$.
In classical growth mechanics the Cauchy stress is written as
\begin{equation}
	\boldsymbol{\sigma}
    =\Je^{-1}\mathbf{g}^\sharp\frac{\partial \We_0}{\partial \Fe}\Fe^\star\,.
\end{equation}
Therefore,
\begin{equation}\label{Cauchy-Stress-Classiical}
	\boldsymbol{\sigma}
    =\Je^{-1}\mathbf{g}^\sharp \Jg^{-1} \frac{\partial \We}{\partial \Fe}\Fe^\star
    =J^{-1}\mathbf{g}^\sharp  \frac{\partial \We}{\partial \Fe}\Fe^\star\,.
\end{equation}
From \eqref{Elastic-Energy}, one can write
\begin{equation}
	\frac{\partial \We}{\partial \Fe}=\frac{\partial \We}{\partial \mathbf{F}}\frac{\partial \mathbf{F}}{\partial \Fe} =\frac{\partial \We}{\partial \mathbf{F}}\Fg^\star
    \,.
\end{equation}
Thus,
\begin{equation}
	\frac{\partial W}{\partial \Fe}\,\Fe^{\star}
    =\frac{\partial \We}{\partial \mathbf{F}}\Fg^\star\,\Fe^{\star}
    =\frac{\partial \We}{\partial \mathbf{F}}(\Fe\Fg)^\star
    = \frac{\partial \We}{\partial \mathbf{F}}\mathbf{F}^\star
    \,.
\end{equation}
Therefore, from \eqref{Cauchy-Stress-Classiical} we have
\begin{equation}
	\boldsymbol{\sigma}
    =J^{-1}\mathbf{g}^\sharp  \frac{\partial \We}{\partial \Fe}\Fe^\star
    =J^{-1}\mathbf{g}^\sharp \frac{\partial \We}{\partial \mathbf{F}}\mathbf{F}^\star
    \,.
\end{equation}
This implies that the Cauchy stresses calculated using the two approaches are identical. 
\end{remark}

Arteries are effectively monoclinic solids.
A monoclinic solid in its initial state has three material preferred directions $\mathring{\mathbf{N}}_1(X)$, $\mathring{\mathbf{N}}_2(X)$, and $\mathring{\mathbf{N}}_3(X)$ such that $\mathring{\mathbf{N}}_1\cdot\mathring{\mathbf{N}}_2\neq 0$ and $\mathring{\mathbf{N}}_3$ is normal to the plane of $\mathring{\mathbf{N}}_1$ and $\mathring{\mathbf{N}}_2$ \citep{Merodio2020}. 
In the material manifold the three vectors representing the material preferred directions are $\mathbf{N}=\Fg^*\mathring{\mathbf{N}}_i=\Fg^{-1}\mathring{\mathbf{N}}_i$, $i=1,2,3$.
A monoclinic solid has an integrity basis $\{I_1,\hdots,I_9\}$, where \citep{Spencer1986}
\begin{equation} \label{I1-7-Definitions}
\begin{aligned}
	& I_1=\mathrm{tr}\,\mathbf{C}=C^A{}_A\,,&& 
	I_2=\mathrm{det}\,\mathbf{C}~\mathrm{tr}_{\mathbf{G}}\mathbf{C}^{-1}
	=\mathrm{det}(C^A{}_B)(C^{-1})^D{}_D\,,\quad 
	I_3=\mathrm{det}\mathbf{C}=\mathrm{det}(C^A{}_B)\,, \\
	& I_4=\mathbf{N}\cdot\mathbf{C}\cdot\mathbf{N}_1=N^A_1N^B_1\,C_{AB}\,,&& 
	I_5=\mathbf{N}_1\cdot\mathbf{C}^2\cdot\mathbf{N}_1
	=\mathbf{N}_1\cdot\mathbf{C}^{\flat}\mathbf{G}^{\sharp}\mathbf{C}^{\flat}\cdot\mathbf{N}_1
	=N^A_1N^B_1\,C_{BM}\,C^M{}_A\,, \\
	& I_6=\mathbf{N}_2\cdot\mathbf{C}\cdot\mathbf{N}_2=N^A_2N^B_2\,C_{AB}\,,&& 
	I_7=\mathbf{N}_2\cdot\mathbf{C}^2\cdot\mathbf{N}_2
	=\mathbf{N}_2\cdot\mathbf{C}^{\flat}\mathbf{G}^{\sharp}\mathbf{C}^{\flat}\cdot\mathbf{N}_2
	=N^A_2N^B_2\,C_{BM}\,C^M{}_A\,, \\
	& I_8=\mathcal{I}\,\mathbf{N}_1\cdot\mathbf{C}\cdot \mathbf{N}_2\,,&&
	I_9=\mathcal{I}^2\,,\qquad \mathcal{I}=\mathbf{N}_1\cdot \mathbf{N}_2\,.
\end{aligned}
\end{equation}
The elastic energy function of a monoclinic solid depends on these nine invariants, i.e.,
\begin{equation}
	\We = \We(X,I_1,I_2,I_3,I_4,I_5,I_6,I_7,I_8,I_9) \,.
\end{equation}
In terms of the integrity basis, the Cauchy stress has the following representation 
\begin{equation}
\begin{aligned}
	\boldsymbol{\sigma} 
	&= \frac{2}{\sqrt{I_3}} \Big\{ \left(I_2\,W_2+I_3\,W_3\right)\mathbf{g}^{\sharp} 
	+W_1\,\mathbf{b}^{\sharp}
	-I_3\,W_2\,\mathbf{c}^{\sharp}
	+W_4\,\mathbf{n}_1\otimes\mathbf{n}_1
	+W_5\left[\mathbf{n}_1\otimes(\mathbf{b}^{\sharp}\mathbf{g}\mathbf{n}_1)
	+(\mathbf{b}^{\sharp}\mathbf{g}\mathbf{n}_1)\otimes\mathbf{n}_1 \right] \\
	& \qquad +W_6\,\mathbf{n}_2\otimes\mathbf{n}_2
	+W_7\left[\mathbf{n}_2\otimes(\mathbf{b}^{\sharp}\mathbf{g}\mathbf{n}_2)
	+(\mathbf{b}^{\sharp}\mathbf{g}\mathbf{n}_2)\otimes\mathbf{n}_2 \right]
	+\mathcal{I}\,W_8\left(\mathbf{n}_1\otimes\mathbf{n}_2+\mathbf{n}_2\otimes\mathbf{n}_1\right) 
	\Big\} \,,
\end{aligned}
\end{equation}
where $\mathbf{n}_j=\mathbf{F}\mathbf{N}_j$ ($j=1,2$), $W_i=\dfrac{\partial W}{\partial I_i}$, $i=1,\hdots,9$, and $W$ is the total energy, which can have non-elastic contributions, e.g., when a growth energy $\Wg$ is considered $W=\We+\Wg$.

For an incompressible solid $I_3=1$, $W_3=0$, and the Cauchy stress includes an indeterminate part $-p\,\mathbf{g}^{\sharp}$. In this case, the Cauchy stress admits the following representation:
\begin{equation} \label{CauchyStress-Monoclinic}
\begin{aligned}
	\boldsymbol{\sigma} 
	&= (-p+2I_2\,W_2)\,\mathbf{g}^{\sharp} 
	+2W_1\,\mathbf{b}^{\sharp}
	-2W_2\,\mathbf{c}^{\sharp}
	+2W_4\,\mathbf{n}_1\otimes\mathbf{n}_1
	+2W_5\left[\mathbf{n}_1\otimes(\mathbf{b}^{\sharp}\mathbf{g}\mathbf{n}_1)
	+(\mathbf{b}^{\sharp}\mathbf{g}\mathbf{n}_1)\otimes\mathbf{n}_1 \right] \\
	& \qquad +2W_6\,\mathbf{n}_2\otimes\mathbf{n}_2
	+2W_7\left[\mathbf{n}_2\otimes(\mathbf{b}^{\sharp}\mathbf{g}\mathbf{n}_2)
	+(\mathbf{b}^{\sharp}\mathbf{g}\mathbf{n}_2)\otimes\mathbf{n}_2 \right]
	+2\mathcal{I}\,W_8\left(\mathbf{n}_1\otimes\mathbf{n}_2+\mathbf{n}_2\otimes\mathbf{n}_1\right) 
	\,.
\end{aligned}
\end{equation}

\begin{remark}
In a pure elasticity problem, one can replace $-p + 2I_2 W_2$ with $-p$, since $p$ is an unknown determined as part of the solution. However, in a growth mechanics problem, $p$ appears explicitly in the growth equation \eqref{Growth-Equation-Metric-Incompressible}, and one must either retain the full expression $-p + 2I_2 W_2$ or replace $-p$ in the growth equation with $-p - 2I_2 W_2$.
\end{remark}

\subsubsection{Dissipation potential}

Bulk growth is a dissipative process. This means that in any mechanical formulation of growth, dissipation due to the evolution of the growth tensor must be taken into account. Let us assume the existence of a dissipation potential (or Rayleigh dissipation function) $\phi=\phi(X,\mathbf{F},\Fg,\dot{\Fg},\boldsymbol{G},\boldsymbol{g})$. Objectivity implies that  $\phi=\hat{\phi}(X,\mathbf{C}^{\flat},\Fg,\dot{\Fg},\boldsymbol{G})$.
Let us assume that $\phi$ is a convex function of $\dot{\Fg}$ \citep{ziegler1958attempt,ziegler1987derivation,Germain1983,Goldstein2002,Kumar2016}. 
A part of the generalized force that corresponds to the evolution of growth tensor is related to the dissipation potential as
\begin{equation}
	\boldsymbol{B}_g^d =- \frac{\partial \phi}{\partial \dot{\Fg}}\,.
\end{equation}
\subsubsection{Growth energy} \label{Sec:RemodelingEnergy}

Drawing on the variational framework introduced by \cite{Francfort98} to model crack evolution, \cite{KumarYavari2023} proposed a class of evolution energies that represent the energetic expenditure associated with cellular growth and remodeling. These evolution energies are postulated to compete with the elastic strain energy, and this competition is assumed to govern the overall growth and remodeling process.
They applied this approach in the previous work to the case of remodeling of fiber-reinforced materials.
Motivated by this study, we propose to introduce a \emph{growth energy}  which controls the tendency of the material to grow in response to changes in mechanical loading.
We denote it as $\Wg$ with the following functional form
\begin{equation} \label{Growth-Energy}
	\Wg=\Wg(X,\mathbf{C}^{\flat},\mathbf{S},\nabla^{\mathbf{G}}\Fg,\mathbf{G})\,,
\end{equation}
where $\nabla^{\mathbf{G}}\Fg$ is the covariant derivative of the growth tensor in the reference configuration and has the following components
\begin{equation}
	\cFg^A{}_{B|C}=\frac{\partial \cFg^A{}_B}{\partial X^C}+\Gamma^A{}_{CD}  \cFg^D{}_B
	-\Gamma^D{}_{BC}  \cFg^A{}_D
	\,.
\end{equation}
It should be noted that there is no reason to expect that growth energy explicitly depends only on strain; it may depend on stress as well as emphasized in \eqref{Growth-Energy}. 
Instead of postulating a growth equation (or an evolution law) \citep{Goriely2017}, our idea is to derive all the governing equations, including a growth equation, using a single variational principle.

\begin{remark}
It should be emphasized that \eqref{Growth-Energy} is a constitutive choice. Another choice is the followig
\begin{equation}
	\Wg=W_g(X,\mathbf{C}^{\flat},\Fg,\nabla^{\mathring{\mathbf{G}}}\Fg,\mathring{\mathbf{G}})
	=W_g(X,\mathbf{C}^{\flat},\Fg^*\Fg,\Fg^*\nabla^{\mathring{\mathbf{G}}}\Fg,\Fg^*\mathring{\mathbf{G}})
	=\Wg(X,\mathbf{C}^{\flat},\Fg^*\nabla^{\mathring{\mathbf{G}}}\Fg,\mathbf{G})
	\,.
\end{equation}
\end{remark}

\subsection{Balance Laws} \label{Sec:BalanceLaws}

In this section, the governing equations of growing bodies are derived in a variational setting. In addition to the standard governing equations of nonlinear elasticity, a \emph{growth equation} is derived.

\subsubsection{Balance of mass}

The mass density at the material point $X\in\mathcal{B}$ at time $t$ is denoted by $\rho_0(X,t)$. 
The balance of mass for a growing body in integral form for any subbody $\mathcal{U} \subset \mathcal{B}$ is written as
\begin{equation}\label{Mass-Balance}
	\frac{d}{dt} \int_{\mathcal U} \rho_0\,dV = \int_{\mathcal U} S_m \,dV\,,
\end{equation}
where $S_m=S_m(X,t)$ is the material rate of change of mass per unit (stress-free) volume. 
Knowing that $dV=\sqrt{\det\mathbf G}\,dX^1 \wedge dX^2 \wedge dX^3$ and $\mathbf{G} = \Fg^*\mathring{\mathbf{G}}\,$, from \eqref{Mass-Balance} one concludes that $\rho_0=\rho_0(X,\Fg_t(X),\mathring{\mathbf{G}}(X),t)$.
Using the identity $\dfrac{d}{d t}\,\det\mathbf{G}= \mathbf{G}^{-1}\!:\!\dot{\mathbf{G}}\,\det\mathbf{G}$, we have $\dfrac{d}{dt}dV=\frac{1}{2}\mathbf{G}^{-1}\!:\!\dot{\mathbf{G}}\,dV$. Thus
\begin{equation}
	\frac{d}{dt} \int_{\mathcal U} \rho_0\,dV 
	=\int_{\mathcal U} \left(\dot{\rho}_0 +\rho_0 \frac{1}{2}\mathbf{G}^{-1}\!:\!\dot{\mathbf{G}}\right)dV
	= \int_{\mathcal U} S_m \,dV\,.
\end{equation}
Therefore, the local form of the balance of mass is written as
\begin{equation} \label{Mass-Balance-Local}
	\dot{\rho}_0 + \frac{1}{2}\rho_0\,\operatorname{tr}\dot{\mathbf G} = S_m \,.
\end{equation}
From $\mathbf{G}=\Fg^{\star}\,\mathring{\mathbf{G}}\,\Fg$ with $\mathring{\mathbf{G}}$ time-independent, we get
$\dot{\mathbf{G}}=\dot{\Fg}^{\star}\,\mathring{\mathbf{G}}\,\Fg+\Fg^{\star}\,\mathring{\mathbf{G}}\,\dot{\Fg}$. Note that 
$\operatorname{tr}\dot{\mathbf G} =\dot{\mathbf{G}}:\mathbf{G}^{-1}=\operatorname{tr}(\dot{\mathbf{G}}\,\mathbf{G}^{-1})$.
In components, $G_{AB}=\mathring{G}_{MN}\,\cFg^{M}{}_{A}\,\cFg^{N}{}_{B}$, and hence
$\dot{G}_{AB}=\mathring{G}_{MN}\big(\dot{\cFg}^{M}{}_{A}\,\cFg^{N}{}_{B}+\cFg^{M}{}_{A}\,\dot{\cFg}^{N}{}_{B}\big)$. Thus,
$\dot{G}_{AB}\,G^{BA}=2\,\dot{\cFg}^{M}{}_{A}\,\cFg^{\,-A}{}_{M}=2\,\operatorname{tr}(\dot{\Fg}\,\Fg^{-1})$. Therefore, $\operatorname{tr}\dot{\mathbf G}=2\,\operatorname{tr}(\dot{\Fg}\,\Fg^{-1})$.
Now balance of mass is rewritten as
\begin{equation} \label{Mass-Balance-Local-Fg}
	\dot{\rho}_0 + \rho_0\,\operatorname{tr}(\dot{\Fg}\,\Fg^{-1}) = S_m \,.
\end{equation}
From $dV=\sqrt{\det\mathbf G}\,dX^1 \wedge dX^2 \wedge dX^3$ and $\mathbf{G} =\Fg^\star\mathring{\mathbf{G}}\Fg$, one can see that $dV=\det\Fg\,\sqrt{\det\mathring{\mathbf G}}\,dX^1 \wedge dX^2 \wedge dX^3=\det\Fg \,d\mathring{V}$.
The growth Jacobian is defined such that $dV=\Jg\,d\mathring{V}$. Thus, $\Jg=\det\Fg$, and hence, $\dot{\Jg}=\det\Fg\,\Fg^{-1}\!:\!\dot{\Fg}=\Jg\,\Fg^{-1}\!:\!\dot{\Fg}$. Thus, the balance of mass is rewritten as: 
\begin{equation}
    \dot{\rho}_0 + \rho_0\,\dfrac{\dot{\Jg}}{\Jg}= S_m
    \,.
\end{equation}

\subsubsection{The Lagrange-d'Alembert principle} \label{Sec:LD-Principle}

In this section we derive the governing equations of a body undergoing finite deformations while growing using the Lagrange-d'Alembert principle. For bulk growth, one has the two independent variations $(\delta\varphi,\delta\Fg)$.
The Lagrangian density is defined as $\mathcal{L} = \mathcal{T} - \rho_0W$, where $\mathcal{T} = \frac{1}{2}\rho_0\Vert\boldsymbol V\Vert^2_{\boldsymbol g}= \frac{1}{2}\rho_0 \llangle \mathbf{V},\mathbf{V}\rrangle_{\boldsymbol g}$ is the kinetic energy density, and $W=\We+\Wg$ with $\We$ and $\Wg$ being the elastic and growth energies, respectively.
The Lagrange-d’Alembert variational principle states that the physical configuration of the growing body satisfies the following identity \citep{Lanczos1962, MarsRat2013}:
\begin{equation} \label{LD-Principle}
\begin{aligned}
	& \delta \int_{t_1}^{t_2}  \int_{\mathcal B} \mathcal{L} \, dV \, \mathrm dt +
	\int_{t_1}^{t_2} \int_{\mathcal B} 
	\boldsymbol{B}_g\!:\!\delta\Fg  \, dV \, \mathrm dt +
	\int_{t_1}^{t_2} \int_{\mathcal B} \rho_0 \llangle \boldsymbol{B} , 
	\delta\varphi \rrangle_{\boldsymbol{g}} \, dV\, \mathrm dt 
	+ \int_{t_0}^{t_1} \int_{\mathcal B}  \llangle S_m \mathbf{V}, 
	\delta\varphi \rrangle_{\mathbf{g}}  \, \mathrm{d}V \, \mathrm{d}t \\
	& \qquad +\int_{t_1}^{t_2} \int_{\partial \mathcal B} \llangle \boldsymbol{T} , 
	\delta\varphi \rrangle_{\boldsymbol{g}} \, dA \,\mathrm dt  = 0 \,,
\end{aligned}
\end{equation}
for any variation fields $\delta\varphi$ and $\delta\Fg$,\footnote{It is assumed that $\delta\varphi(X,t_1)=\delta\varphi(X,t_2)=0$, and $\delta\Fg(X,t_1)=\delta\Fg(X,t_2)=\mathbf{0}$.} where $\boldsymbol{B}$ and $\boldsymbol{T}$ are, respectively, the body force per unit mass and the boundary traction per unit undeformed area. Note that $S_m \mathbf{V}$ is the rate of momentum corresponding to mass growth (resorption).
The growth generalized force $\boldsymbol{B}_g$ is assumed to be the sum of a dissipative force $\boldsymbol{B}_g^d$ (corresponding to a dissipation potential) and a non-dissipative force $\hat{\boldsymbol{B}}_g$ (a growth configurational body force \citep{KRLP18}) that depends on both stress and growth distortion:
\begin{equation}
	\boldsymbol{B}_g =- \frac{\partial \phi}{\partial \dot{\Fg}}
	+\hat{\boldsymbol{B}}_g(\Fg,\mathbf{S},\mathbf{G})\,.
\end{equation}
Next, we carry out the variations.

\begin{itemize}[topsep=0pt,noitemsep, leftmargin=10pt]
\item $\delta\varphi$ \textbf{variations:}
It is straightforward to show that this variation yields the balance of linear momentum along with Neumann boundary conditions \citep{KumarYavari2023}:
\begin{equation} \label{EL-Compressible}
\begin{dcases}
	\operatorname{Div}\left(\rho_0\mathbf{g}^{\sharp}\frac{\partial W}{\partial \mathbf{F}}\right)
	+\rho_0\mathbf{B}=\rho_0\mathbf{A}\,, & \text{in~}\mathcal{B}\,,\\
	 \rho_0 \mathbf{g}^{\sharp}\frac{\partial W}{\partial \mathbf{F}}\hat{\mathbf{N}} =\mathbf{T}\,, 
	 & \text{on~} \partial_N \mathcal{B}\,.
\end{dcases}
\end{equation}
If the growing material is incompressible, a term $p(J-1)$ is added to the Lagrangian density. Thus, $\delta\mathcal{L}=\delta\mathcal{T}-\delta (\rho_0 W)+p\,\delta J=\delta\mathcal{T}-\delta (\rho_0 W)+pJ\,\mathbf{F}^{-1}\!:\!\delta \mathbf{F}$. The Euler-Lagrange equations and the natural boundary conditions are written as
\begin{equation} \label{EL-Incompressible}
\begin{dcases}
	\operatorname{Div}\left[-pJ\,\mathbf{F}^{-1}+\rho_0 \mathbf{g}^{\sharp}
	\frac{\partial W}{\partial \mathbf{F}}\right]
	+\rho_0\mathbf{B}=\rho_0\mathbf{A}\,, &\text{in~}\mathcal{B}\,,\\
	\left[-pJ\,\mathbf{F}^{-1}+\rho_0 \mathbf{g}^{\sharp}\frac{\partial W}{\partial \mathbf{F}}\right]\hat{\mathbf{N}} 
	=\mathbf{T}\,, & \text{on~} \partial_N \mathcal{B}\,.
\end{dcases}
\end{equation}

\item $\delta\Fg$ \textbf{variations:}
Recall that $\mathcal{T} = \frac{1}{2}\rho_0 \llangle \mathbf{V},\mathbf{V}\rrangle_{\boldsymbol g}$, and hence $\delta\mathcal{T} = \frac{1}{2} \delta\rho_0 \llangle \mathbf{V},\mathbf{V}\rrangle_{\boldsymbol g}$.
One can show that $\delta\rho_0=-\frac{1}{2}\rho_0\operatorname{tr}(\delta\mathbf{G})=-\frac{1}{2}\rho_0\mathbf{G}^{-1}\!:\!\delta\mathbf{G}$ \citep{Yavari2010}.
Thus
\begin{equation} 
	\delta (\mathcal{T}\,dV)  = -\frac{1}{4} \rho_0 \llangle \mathbf{V},\mathbf{V}\rrangle_{\boldsymbol g} \,
	\mathbf{G}^{-1}\!:\!\delta\mathbf{G}\, dV
	+\frac{1}{2} \rho_0 \llangle \mathbf{V},\mathbf{V}\rrangle_{\boldsymbol g} 
	\left(\frac{1}{2} \operatorname{tr}(\delta\mathbf{G}) dV\right)
	= 0
	 \,,
\end{equation}
where use was made of the identity $\delta dV=\frac{1}{2} \operatorname{tr}(\delta\mathbf{G}) dV$.
 Variation of the total energy is written as
\begin{equation} 
	\delta (\rho_0 W dV)=\rho_0 \frac{\partial W}{\partial \mathbf{G}}\!:\!\delta\mathbf{G}
	 \,,
\end{equation}
where $W=\We+\Wg$ and use was made of the fact that $\delta (\rho_0 dV)=0$.
Thus
\begin{equation} 
	\delta (\rho_0 W)
	= \rho_0 \frac{\partial W}{\partial \mathbf{G}} \!:\!\delta\mathbf{G}
	 \,.
\end{equation}

For anisotropic solids structural tensors are included in the arguments of the energy and the above variation should be modified to read
\begin{equation} 
	\delta W= \left[
	\frac{\partial \We}{\partial \mathbf{G}}+\frac{\partial \Wg}{\partial \mathbf{G}}
	\right]
	\!:\!\delta\mathbf{G}
	+\left[\frac{\partial \We}{\partial \boldsymbol{\Lambda}}
	+\frac{\partial \Wg}{\partial \boldsymbol{\Lambda}}\right]
	\!:\!\delta\boldsymbol{\Lambda}
	 \,,
\end{equation}
where $\boldsymbol{\Lambda}$ is a collection of structural tensors that explicitly depends on the symmetry group. For now, let us ignore the structural tensors.
From \eqref{Material-Metric}, one writes $\delta\mathbf{G}=\delta(\Fg^{\star}\mathring{\mathbf{G}}\Fg)=(\delta\Fg)^{\star}\mathring{\mathbf{G}}\Fg+\Fg^{\star}\mathring{\mathbf{G}}\delta\Fg$.
Thus
\begin{equation} 
	\left[\frac{\partial \We}{\partial \mathbf{G}}+\frac{\partial \Wg}{\partial \mathbf{G}}\right]\!:\!
	\left[ 
	(\delta\Fg)^{\star}\mathring{\mathbf{G}}\Fg+\Fg^{\star}\mathring{\mathbf{G}}\delta\Fg 
	\right]
	=2\mathring{\mathbf{G}}\Fg^{\star}
	\left[\frac{\partial \We}{\partial \mathbf{G}}+\frac{\partial \Wg}{\partial \mathbf{G}}\right]
	\!:\!\delta\Fg	 
	=2\Fg^{-\star}\mathbf{G} 
	\left[\frac{\partial \We}{\partial \mathbf{G}}+\frac{\partial \Wg}{\partial \mathbf{G}}\right]\!:\!\delta\Fg	 
	\,.
\end{equation}
Therefore, the variational principle \eqref{LD-Principle} is simplified to read
\begin{equation} 
	\int_{t_1}^{t_2}  \int_{\mathcal B} 
	\left\{ -2\rho_0\Fg^{-\star}\mathbf{G} 
	\left[\frac{\partial \We}{\partial \mathbf{G}}+\frac{\partial \Wg}{\partial \mathbf{G}}\right]
	- \frac{\partial \phi}{\partial \dot{\Fg}}+\hat{\boldsymbol{B}}_g \right\}\!:\delta\Fg \, dV \,\mathrm dt = 0 \,,
\end{equation}
where the identity $\mathbf{G}^{-1}\!:\!\delta\mathbf{G}=2\Fg^{-1}\!:\!\delta\Fg$ was used in the second term on the right-hand side.
This implies that the \emph{growth equation} for isotropic solids reads 
\begin{equation} \label{Growth-Equation-Metric}
	\frac{\partial \phi}{\partial \dot{\Fg}}=-2\rho_0\Fg^{-\star}\mathbf{G} 
	\left[\frac{\partial \We}{\partial \mathbf{G}}+\frac{\partial \Wg}{\partial \mathbf{G}}\right]
	+\hat{\boldsymbol{B}}_g
	\,.
\end{equation}

\end{itemize}

\begin{remark}
When gradient effects are considered in the growth energy, one can show that an extra term will appear in the growth equation:
\begin{equation} \label{Growth-Equation-Metric-Gradient}
	\frac{\partial \phi}{\partial \dot{\Fg}}=-2\rho_0\Fg^{-\star}\mathbf{G} 
	\left[\frac{\partial \We}{\partial \mathbf{G}}+\frac{\partial \Wg}{\partial \mathbf{G}}\right]
	-\operatorname{Div} 
	\left(\frac{\partial \Wg}{\partial \nabla^{\mathbf{G}}\Fg}\right)
	+\hat{\boldsymbol{B}}_g
	\,.
\end{equation}
There is also a new natural boundary condition:
\begin{equation} 
	\frac{\partial \Wg}{\partial \nabla^{\mathbf{G}}\Fg}\cdot\mathbf{N}
	=\mathbf{0}\,,\qquad \text{on~} \partial \mathcal{B}
	 \,.
\end{equation}
\end{remark}

\begin{remark}
When the material is incompressible, one adds a term $p(J-1)$ to the Lagrangian density. When $\varphi$ varies, the Lagrangian variation is calculated as $\delta\mathcal{L}=\delta\mathcal{T}-\delta (\rho_0W)+p\,\delta J=\delta\mathcal{T}-\delta (\rho_0W)+pJ\,\mathbf{F}^{-1}\!:\!\delta \mathbf{F}$.
Recall that $J=(\det\mathbf{g})^{\frac{1}{2}}(\det\mathbf{G})^{-\frac{1}{2}}\det\mathbf{F}$, and hence when $\Fg$ varies, there is a corresponding $\delta J=-\frac{1}{2}J\,(\det\mathbf{G})^{-1}\,\delta(\det\mathbf{G})=-\frac{1}{2}J\,\mathbf{G}^{-1}\!:\!\delta\mathbf{G}=-J\,\Fg^{-1}\!:\!\delta\Fg$. Thus, for $\Fg$-variations: $\delta\mathcal{L}=-\delta (\rho_0W)+p\,\delta J=-\delta (\rho_0W)-pJ\,\Fg^{-1}\!:\!\delta\Fg$.
In this case the growth equation is modified to read
\begin{equation} \label{Growth-Equation-Metric-Incompressible}
	\frac{\partial \phi}{\partial \dot{\Fg}}=-2\rho_0\Fg^{-\star}\mathbf{G} 
	\left[\frac{\partial \We}{\partial \mathbf{G}}+\frac{\partial \Wg}{\partial \mathbf{G}}\right]
	-pJ\,\Fg^{-1}
	\,.
\end{equation}
\end{remark}
\begin{remark}
Our formulation exhibits a close connection with gradient flow models recently proposed for morphoelasticity \citep{Ouzeri2026}. In particular, the evolution equation for the growth tensor derived in this work may be interpreted as a dissipative flow in the space of internal metrics, where the driving force is the Eshelby-type configurational stress and the dissipation potential defines the effective metric structure on this space. This interpretation is consistent with the view that growth and remodeling correspond to the steepest descent evolution of an energy functional under a prescribed dissipation mechanism, as in the gradient structure of isotropic morphoelastic bodies.
\end{remark}

\subsubsection{Isotropic growth: Material metric and the growth equation}

Let us consider cylindrical coordinates $(R,\Theta,Z)$ in the reference configuration of an artery. In its reference configuration, the artery is assumed to have inner and outer radii $R_i$ and $R_o$, respectively.
The artery is assumed to be made of an incompressible isotropic solid reinforced by two families of helical fibers. In the reference configuration the unit tangent vector to the two fibers at $X=(R,\Theta,Z)\in\mathcal{B}$ are denoted by $\mathbf{N}_1(X)$ and $\mathbf{N}_2(X)$, which are $\mathbf{G}$-unit vectors. 
For helical fibers $\mathbf{N}_1(X)=\mathbf{N}_1(R)$ and $\mathbf{N}_2(X)=\mathbf{N}_2(R)$. The distributed fibers make the artery effectively a monoclinic solid. The tangent vectors to the deformed fibers in the current configuration are denoted as $\mathbf{n}_1=\mathbf{F}\mathbf{N}_1$ and $\mathbf{n}_2=\mathbf{F}\mathbf{N}_2$. We assume isotropic growth $\Fg=\mathfrak{g}\,\mathbf{I}$.




We consider a dissipation potential $\phi=\phi(\mathfrak{g},\dot{\mathfrak{g}},I_1,\hdots,I_9)$, which is convex in the rate of growth factors. For our numerical examples we use a quadratic dissipation potential of the following form
\begin{equation} \label{Anisotropic-Dissipation}
	\phi=
	\frac{1}{2}K\,\dot{\mathfrak{g}}^2
	\,,
\end{equation}
where $K=K(I_1,\hdots,I_9)>0$.
The generalized force corresponding to the evolution of the growth factor is related to the dissipation potential as
\begin{equation} \label{Anisotropic-Growth-Forces}
	B_{g} = -\frac{\partial \phi}{\partial \dot{\mathfrak{g}}_R} + \hat{B}_g
	=-K\,\dot{\mathfrak{g}} + \hat{B}_g	\,.
\end{equation}

For our numerical examples, we consider a growth energy that has the following form
\begin{equation} \label{Anisotropic-Growth-Energy-Gradient}
	\Wg(\mathfrak{g},,I_1,\hdots,I_9)
	=\frac{1}{2}\kappa_{g}\,(\mathfrak{g}-1)^2
	+\frac{1}{2}\hat{\kappa}_{g}\,\|\nabla\mathfrak{g}\|^2
	\,.
\end{equation}
where $\kappa_{g}=\kappa_{g}(I_1,\hdots,I_9)>0$.

\subsection{The first law of thermodynamics}

The first law of thermodynamics (the balance of energy) is expressed as \citep{EpsteinMaugin2000,LubardaHoger2002}:\footnote{Note that the last term in \eqref{eq:Thermo_First} is included to account for changes in the internal and kinetic energies of the system resulting from bulk growth characterized by the material rate of change of mass $S_m$. In other words, a growing body is an open system, and the last term on the right-hand side explicitly accounts for this. We should also note that the material volume element is explicitly time dependent. One can still use an energy per unit volume. In that case instead of $S_m W$ on the right-hand side one would have $\frac{1}{2}\mathbf{G}^{-1}\!:\!\dot{\mathbf{G}} \,W$ instead.}
\begin{equation}\label{eq:Thermo_First}
\begin{aligned}
	\frac{d}{dt}\int_{\mathcal{U}} \rho_0\left(W
	+\frac{1}{2} \llangle \mathbf{V},\mathbf{V}\rrangle_{\boldsymbol g} \right)dV
	& =\int_{\mathcal{U}}\rho_0 \left(\llangle\mathbf{B},\mathbf{V}\rrangle_{\mathbf{g}}
	+R\right)dV
	+\int_{\partial \mathcal{U}}\left(\llangle \mathbf{T},\mathbf{V}\rrangle_{\mathbf{g}}
	+H\right)dA \\
	& \quad + \int_{\mathcal{U}} \hat{\mathbf{B}}_g\!:\! \dot{\Fg} dV +\int_{\mathcal{U}} 
	S_m\left(W+\frac{1}{2}  \llangle \mathbf{V},\mathbf{V}\rrangle_{\boldsymbol g} \right)
	 dV\,.
\end{aligned}
\end{equation}
Here, $\mathcal{U} \subset \mathcal{B}$ is an arbitrary sub-body, $W$ denotes the energy function or internal energy density per unit mass in the reference configuration, $R = R(X, t)$ is the heat supply per unit mass, $H = -\llangle \mathbf{Q}, \hat{\mathbf{N}} \rrangle_{\mathbf{G}}$ represents the heat flux, $\mathbf{Q} = \mathbf{Q}(X, T, \mathrm{d}T, \mathbf{C}, \mathbf{G})$ is the external heat flux per unit area, $\hat{\mathbf{N}}$ is the $\mathbf{G}$-unit normal to the boundary $\partial\mathcal{B}$, $T = T(X, t)$ is the absolute temperature field, $\mathrm{d}T$ is the exterior derivative of temperature (a $1$-form), and ${S_m=S_m(X,t)}$ is the material rate of change of mass per unit (stress-free) volume. Note that $S_m$ is identically zero in the absence of bulk growth.


The local form of the balance of energy is written as
\begin{equation} 
\begin{aligned}
	\rho_0 \dot{W} &= \rho_0 R+\mathbf{P}\!:\!\nabla\mathbf{V}- \operatorname{Div} \mathbf{Q}
	+\llangle \operatorname{Div}\mathbf{P}
	+ \rho_0(\mathbf{B} - \mathbf{A}),\mathbf{V} \rrangle_{\mathbf{g}}
	+ \hat{\mathbf{B}}_g\!:\! \dot{\Fg}\\& \quad+ \left(S_m-\dot{\rho}_0 - \frac{1}{2}\rho_0\,\dot{\mathbf G}\!:\!\mathbf G^{-1} \right)
	\left(W+\frac{1}{2} \llangle \mathbf{V},\mathbf{V}\rrangle_{\boldsymbol g} \right)
	\,.
\end{aligned}
\end{equation}
At this stage, we have not yet established that the first Piola-Kirchhoff stress is expressed as $\mathbf{P} = \dfrac{\partial W}{\partial \mathbf{F}}$, as this relationship is derived after using the second law of thermodynamics (see \citet{SadikYavari2025} for more details). However, to facilitate the calculations, we will assume it for now.
We also use the local balance of mass \eqref{Mass-Balance-Local}.
It can be readily shown that $\mathbf{P} \!:\! \nabla \mathbf{V} = \frac{1}{2} \mathbf{S} \!:\! \dot{\mathbf{C}}^{\flat}$.
Consequently, the local form of the energy balance becomes:
\begin{equation} \label{Local-FirstLaw}
	\rho_0 \dot{W} = \rho_0 R+\frac{1}{2}\mathbf{S}\!:\! \dot{\mathbf{C}}^{\flat}
	- \operatorname{Div} \mathbf{Q} + \hat{\mathbf{B}}_g\!:\! \dot{\Fg}
	\,.
\end{equation}

\subsection{The second law of thermodynamics} \label{Sec:2ndLaw}

The second law of thermodynamics can be expressed in the form of the material Clausius-Duhem inequality \citep{MarsdenHughes1983,EpsteinMaugin2000,LubardaHoger2002}, which is written as:\footnote{Note that the last term is included to account for changes in the entropy of the system due to bulk growth.}
\begin{equation} 
	\frac{d}{dt}\int_{\mathcal{U}} \rho_0\,\mathcal{N}dV\geq 
	\int_{\mathcal{U}}\rho_0 \frac{R}{T}dV+\int_{\partial\mathcal{U}}\frac{H}{T}dA
	+\int_{\mathcal{U}} S_m\, \mathcal{N} dV
	\,,
\end{equation}
where $\mathcal{N}= \hat{\mathcal N}(X,T, \mathbf{C}^\flat,\mathbf{G})$ is the material entropy density (per unit mas).
The local form of the Clausius-Duhem inequality is expressed as:
\begin{equation} 
	\dot\eta = \rho_0T \dot{\mathcal{N}} - \rho_0 R+ T \operatorname{Div}\left(\frac{\mathbf Q}{T}\right) 
	- \left(S_m-\dot{\rho}_0 - \frac{1}{2}\rho_0\,\dot{\mathbf G}\!:\!\mathbf G^{-1} \right) T \mathcal{N}
	 \geq 0\,,
\end{equation}
where $\dot\eta$ is the rate of energy dissipation. Using the local balance of mass \eqref{Mass-Balance-Local}, this is further reduced to read
\begin{equation} \label{Local-SecondLaw}
	\dot\eta = \rho_0T \dot{\mathcal{N}} - \rho_0 R+ T \operatorname{Div}\left(\frac{\mathbf Q}{T}\right) 
	 \geq 0\,.
\end{equation}
The free energy density is defined as $\Psi = W - T \mathcal{N}$, which can be expressed as $\Psi = \hat{\Psi}(X, T, \mathbf{C}^{\flat}, \mathbf{G})$. Observe that $T \dot{\mathcal{N}} = \dot{W} - \dot{\Psi} - \dot{T} \mathcal{N}$, and consequently:
\begin{equation} 
	\dot{\eta} = \rho_0 \dot{W} -\rho_0\dot{\Psi} - \rho_0\dot{T} \mathcal{N}+ \operatorname{Div}\mathbf{Q}
	- \frac{1}{T} \langle dT, \mathbf{Q} \rangle -\rho_0 R \geq 0\,.
\end{equation}
Substituting \eqref{Local-FirstLaw} into the above inequality, one obtains
\begin{equation} \label{eta-inequality}
	\dot{\eta} = \frac{1}{2}\mathbf{S}\!:\! \dot{\mathbf{C}}^{\flat} + \hat{\mathbf{B}}_g\!:\! \dot{\Fg}	 
	-\rho_0\dot{\Psi} -\rho_0 \dot{T} \mathcal{N}
	- \frac{1}{T} \langle dT, \mathbf{Q} \rangle  \geq 0\,.
\end{equation}
The total time derivative of the free energy is written as
\begin{equation} 
	\dot{\Psi}= \frac{\partial  \hat{\Psi}}{\partial T}\,\dot{T}
	+\frac{\partial  \hat{\Psi}}{\partial \mathbf{C}^{\flat}}\!:\!\dot{\mathbf{C}}^{\flat}
	+\frac{\partial  \hat{\Psi}}{\partial \mathbf{G}}\!:\!\dot{\mathbf{G}}
	= \frac{\partial  \hat{\Psi}}{\partial T}\,\dot{T}
	+\frac{\partial  \hat{\Psi}}{\partial \mathbf{C}^{\flat}}\!:\!\dot{\mathbf{C}}^{\flat}
	+2\Fg^{-\star}\mathbf{G}\frac{\partial \Psi}{\partial \mathbf{G}}\!:\!\dot{\Fg}
	\,.
\end{equation}
Thus, \eqref{eta-inequality} is simplifies as
\begin{equation}
	\dot{\eta} = -\rho_0\left(\mathcal{N} + \frac{\partial  \hat{\Psi}}{\partial T}\right) \dot{T} 
	+\frac{1}{2} \left(\mathbf{S}-2\rho_0\frac{\partial  \hat{\Psi}}{\partial \mathbf{C}^{\flat}} \right)
	\!:\! \dot{\mathbf{C}}^{\flat}
	- \frac{1}{T} \langle dT, \mathbf{Q} \rangle
	+ \left(\hat{\mathbf{B}}_g -2\rho_0\Fg^{-\star}\mathbf{G}\frac{\partial \Psi}{\partial \mathbf{G}}\right) \!:\!\dot{\Fg}  \geq 0\,.
\end{equation}
The above inequality must hold for arbitrary $\dot{T} $, and $\dot{\mathbf{C}}^{\flat}$, and therefore
\begin{equation} \label{second-law-consequences}
	 \mathcal{N}= -  \frac{\partial \hat{\Psi}}{\partial T}\,,\qquad
	 \mathbf{S}=2\rho_0\frac{\partial  \hat{\Psi}}{\partial \mathbf{C}^{\flat}}\,,\qquad
	\dot{\eta} =  - \frac{1}{T} \langle dT, \mathbf{Q} \rangle
	+ \left(\hat{\mathbf{B}}_g -2\rho_0\Fg^{-\star}\mathbf{G}\frac{\partial \Psi}{\partial \mathbf{G}}\right) \!:\!\dot{\Fg}
	\geq 0\,.
\end{equation}
Notice that 
\begin{equation}
	\frac{\partial W}{\partial \mathbf{G}}
	=\frac{\partial W}{\partial \mathbf{G}}\Bigg|_{\mathcal{N},\mathbf{C}^{\flat}}
	=\left[ \frac{\partial \Psi}{\partial \mathbf{G}}+\frac{\partial \Psi}{\partial T}
	\frac{\partial T}{\partial \mathbf{G}}
	 \right]+\frac{\partial T}{\partial \mathbf{G}}\mathcal{N}
	 =\frac{\partial \Psi}{\partial \mathbf{G}}
	\,,
\end{equation}
where the second equality follows from the use of \eqref{second-law-consequences}$_1$.
Using the above identity and the growth equation \eqref{Growth-Equation-Metric} in \eqref{second-law-consequences}$_3$, one obtains\footnote{It turns out that this inequality holds for anisotropic solids as well.}
\begin{equation} 
	\dot{\eta} =  - \frac{1}{T} \langle dT, \mathbf{Q} \rangle + \frac{\partial \phi}{\partial \dot{\Fg}} \!:\!\dot{\Fg}
	  \geq 0\,.
\end{equation}
If an isothermal process is assumed, i.e., $dT=0$, the entropy production is simplified to read
\begin{equation} \label{Entropy-Production-Growth}
	\dot{\eta} =  \frac{\partial \phi}{\partial \dot{\Fg}} \!:\!\dot{\Fg} 	\geq 0\,.
\end{equation}
This is automatically satisfied when $\phi$ is a convex function of $\dot{\Fg}$

\section{Isotropic Bulk Growth of Arteries} \label{Sec:Arteries}

Arteries comprise three layers---intima, media, and adventitia---with the latter two exerting the dominant influence on their mechanical response. In the literature, arteries are often modeled as double-layer shells, ignoring the intima. In our formulation we assume that an artery is radially inhomogeneous, i.e., $\We=\We(R)$. However, in our numerical examples, we will work with a piecewise uniform elastic energy.
The following features have been well established through the analysis of experimental observations for the passive mechanical response and active response through bulk growth in arterial walls:

\begin{itemize}[topsep=2pt,noitemsep, leftmargin=10pt]
\item The inner and outer layers are well represented as a monoclinic solid with two preferred fiber directions or a transversely isotropic solid with one preferred fiber direction \citep{holzapfel2004data}.
\item The contribution of collagen fibers to the elasticity of arterial wall has been observed to be small for low pressures, thus, it is common to describe the elastic response of each layer through an additive decomposition of the strain energy function in the following form: $W = \hat{W}(I_1)+  \tilde{W}(I_4, I_6)$.
\item The artery operates under an oscillating state of stress. Viewing it as a cylindrical shell, it is subjected to internal pressure and longitudinal traction from adjacent tissues.
\item Observations suggest that close to a `homeostatic' axial stretch, no further axial stretch is observed as internal pressure is varied. This observation has been shown to be strongly related to the significant strain stiffening of fibers \citep{Goriely2017}. Use of a Fung-type hyperelastic model allows to capture this behavior.
\item Bulk growth is associated with smooth muscle cell activation in the inner layer, resulting in an increase in wall thickness and a change in the lumen diameter. 
\end{itemize}

Let us consider a hollow circular cylindrical bar of initial length $L$ and inner and outer radii $R_i$ and $R_o$, respectively. We assume that the outer cylindrical boundary is traction-free while the inner boundary is under a time-dependent pressure $p_i(t)$.\footnote{In our analysis, we start with an undeformed artery at time $t=0$. The artery is then loaded by an internal pressure $p_i(t)$ until pressure reaches a value $\mathring{p}_i$ at time $t_i$ and for this pressure the deformed artery has internal radius $\mathring{r}_i$. We assume that in the time interval $[0,t_i)$ there is no growth. Growth starts right at time $t_i$. 
It is well known that arterial growth is accompanied by changes in blood pressure. Depending on the stage of hypertension, systolic blood pressure can increase by approximately $10\%$ to $50\%$ relative to normal levels.
For modeling purposes, one may assume that for $t>t_i$, $p_i(t)$ explicitly depends on the deformed inner radius of the artery, i.e., $p_i = p_i(r(R_i,t)) = p_i(r_i(t))$.
A typical nonlinear relationship between the transmural pressure and the deformed inner radius of the artery is given by
\begin{equation}
	p_i(t) = \mathring{p}_i \, e^{\alpha (r_i(t) - \mathring{r}_i)} \,,
\end{equation}
where $\alpha$ is a material parameter. This form has been used in several arterial wall models; see, e.g., \citep{Langewouters1985,Olufsen2000}. In our numerical examples, we prescribe the internal pressure as a known function of time.} We also assume that the bar has a fixed cylindrically-symmetric distribution of radial and circumferential finite eigenstrains.
The cylindrical coordinates $(R,\Theta,Z)$ are used for the reference configurations.

\subsection{Kinematics} 

We assume that initially and before the growth process starts the bar has an axisymmetric distribution of radial and circumferential eigenstrains. Thus, the material metric is assumed to have the following form \citep{YavariGoriely2013}
\begin{equation}
	\mathring{\mathbf{G}}(R)=
	\begin{bmatrix}
  e^{2\omega_1(R)} & 0  & 0  \\
  0 & R^2 e^{2\omega_2(R)}  & 0  \\
  0 & 0  &  1 
\end{bmatrix}\,,
\end{equation}
where $\omega_1(R)$ and $\omega_2(R)$ are some given functions that quantify the radial and circumferential eigestrains,\footnote{These eigenstrains can have different sources, e.g., defects, temperature changes, swelling, etc.} respectively. Note that $\mathring{\mathbf{G}}$ is non-flat, in general.
We assume that the hollow circular cylindrical bar is reinforced by two families of helical fibers. This makes the artery effectively monoclinic. The unit tangent vectors of the two fiber families before growth are denoted by $\mathring{\mathbf{N}}_1=\mathring{\mathbf{N}}_1(R,\Theta)$, and $\mathring{\mathbf{N}}_2=\mathring{\mathbf{N}}_2(R,\Theta)$. 
These being unit vectors means that $\llangle \mathring{\mathbf{N}}_1,\mathring{\mathbf{N}}_1 \rrangle_{\mathring{\mathbf{G}}}=\llangle \mathring{\mathbf{N}}_2,\mathring{\mathbf{N}}_2 \rrangle_{\mathring{\mathbf{G}}}=1$.
Let $\gamma_1(R)$ and $\gamma_2(R)$ be the angles that $\mathring{\mathbf{N}}_1(R,\Theta)$ and $\mathring{\mathbf{N}}_2(R,\Theta)$ make with $\mathbf{E}_{\Theta}=\frac{\partial}{\partial \Theta}$. Thus
\begin{equation}
	\mathring{\mathbf{N}}_1(R,\Theta)=\frac{\cos\gamma_1(R)}{R \,
	e^{\omega_2(R)}}\,\mathbf{E}_{\Theta}(\Theta)
	+\sin\gamma_1(R)\,\mathbf{E}_Z\,,\qquad
	\mathring{\mathbf{N}}_2(R,\Theta)=\frac{\cos\gamma_2(R)}{R\,
	e^{\omega_2(R)}}\,\mathbf{E}_{\Theta}(\Theta)
	+\sin\gamma_2(R)\,\mathbf{E}_Z\,.
\end{equation}
Let us assume an isotropic growth for which 
\begin{equation} \label{Anisotropic-Growth-Tensor}
	\Fg=
	\begin{bmatrix}
	\mathfrak{g} & 0 & 0 \\
	0 & \mathfrak{g} & 0 \\
	0 & 0 & \mathfrak{g}
	\end{bmatrix}
	\,,
\end{equation}
where $\mathfrak{g}=\mathfrak{g}(R,t)$ is an unknown function to be determined. The initial condition is: $\mathfrak{g}(R,0)=1$.
Recall that $\mathbf{G}=\Fg^*\mathring{\mathbf{G}}=\Fg^\star\mathring{\mathbf{G}}\Fg$, and hence
\begin{equation}  \label{Metric-Anisotropic}
	\mathbf{G}(R,t)= \mathfrak{g}^2(R,t)
	\begin{bmatrix}
	e^{2\omega_1(R)} & 0  & 0  \\
	0 & R^2 \,e^{2\omega_2(R)}  & 0  \\
	0 & 0  &  1
\end{bmatrix}\,.
\end{equation}
In the growing body the unit tangent vectors of the two fiber families are denoted by $\mathbf{N}_1=\mathbf{N}_1(R,\Theta)$, and $\mathbf{N}_2=\mathbf{N}_2(R,\Theta)$ and are represented as follows
\begin{equation}
\begin{aligned}
	\mathbf{N}_1(R,\Theta)
	& =\frac{\cos\gamma_1(R)}{R \,\mathfrak{g}(R,t)\,
	e^{\omega_2(R)}}\,\mathbf{E}_{\Theta}(\Theta)
	+\frac{\sin\gamma_1(R)}{\mathfrak{g}(R,t)}\,\mathbf{E}_Z\,,\\
	\mathbf{N}_2(R,\Theta)
	& =\frac{\cos\gamma_2(R)}{R\,\mathfrak{g}(R,t)\,
	e^{\omega_2(R)}}\,\mathbf{E}_{\Theta}(\Theta)
	+\frac{\sin\gamma_2(R)}{\mathfrak{g}(R,t)}\,\mathbf{E}_Z\,.
\end{aligned}
\end{equation}

For the Euclidean ambient space the cylindrical coordinates $(r,\theta,z)$ are used. The metric of the Euclidean ambient space has the following representation
\begin{equation}
	\mathbf{g}=
	\begin{bmatrix}
  1 & 0  & 0  \\
  0 & r^2  & 0  \\
  0 & 0  &  1 
\end{bmatrix}\,.
\end{equation}
The following deformations are assumed: $(r,\theta,z)=(r(R,t),\Theta,\lambda(t) Z)$.\footnote{These deformations are subsets of Family $3$ universal deformations---deformations that can be maintained in the absence of body forces for any member of a given class of materials \citep{Ericksen1954,Ericksen1955}. See \citet{YavariGoriely2021,YavariGoriely2022} for generalization to anisotropic solids, and specifically, fiber-reinforced solids.} We consider two cases: i) Displacement-control loadings in which $\lambda(t)$ is a given function, and ii) force-control loadings in which $\lambda(t)$ is a function to be determined.
Deformation gradient is written as 
\begin{equation}
	\mathbf{F}=
	\begin{bmatrix}
  r_{,R}(R,t) & 0  & 0  \\
  0 & 1  & 0  \\
  0 & 0  & \lambda(t) 
\end{bmatrix}\,.
\end{equation}

For an incompressible solid, one has
\begin{equation}
    J=\sqrt{\frac{\det\mathbf{g}}{\det\mathbf{G}}}\det\mathbf{F}
    =\frac{\lambda(t)}{R\,\mathfrak{g}^3(R,t)\,
    e^{\omega_1(R)+\omega_2(R)}}\,r(R,t)\,r_{,R}(R,t)=1.
\end{equation}
Thus
\begin{equation} \label{r1-g-relationship}
    r^2(R,t)=r_i^2(t)
    +\frac{2}{\lambda(t)}\int_{R_i}^{R} \xi\,\mathfrak{g}^3(\xi,t)\,e^{\omega_1(\xi)+\omega_2(\xi)}d\xi
    \,,
\end{equation}
where $r_i(t)=r(R_i,t)$ is an unknown function a priori. 


\subsection{Stress} 

The eight invariants (recall that $I_3=1$) are written as
\begin{equation} \label{I1-I9-Definitions}
\begin{aligned}
    I_1 &=  \frac{r^2 e^{-2 \omega_2(R)}}{R^2 \mathfrak{g}^2(R,t)}+\frac{R^2
   \mathfrak{g}^4(R,t) \,e^{2 (\omega_1(R)+\omega_2(R))
   -2 \omega_1(R)}}{\lambda ^2 r^2}+\frac{\lambda^2}{\mathfrak{g}^2(R,t)}
    \,,\\
    I_2 &= \frac{R^2 \mathfrak{g}^2(R,t) \,e^{2 (\omega_1(R)+\omega_2(R))-2
   \omega_1(R)}}{r^2}+\frac{\lambda ^2 r^2 e^{-2 \omega_2(R)}}{R^2
   \mathfrak{g}^4(R,t)}+\frac{\mathfrak{g}^2(R,t) \,e^{2 (\omega_1(R)+\omega_2(R))
   -2 \omega_1(R)-2 \omega_2(R)}}{\lambda^2}  \,\\
   I_4 &= \frac{r^2 \cos^2\gamma_1 \,e^{-2 \omega_2(R)}}{R^2 \mathfrak{g}^2(R,t)}
   +\frac{\lambda^2 \sin^2\gamma_1}{\mathfrak{g}^2(R,t)}\,,\\
   I_5 &= \frac{r^4 \cos^4\gamma_1 \,e^{-4 \omega_2(R)}}{R^4 \mathfrak{g}^4(R,t)}
   +\frac{\lambda^4 \sin^2\gamma_1}{\mathfrak{g}^4(R,t)} \,,\\
   I_6 &= \frac{r^2 \cos^2\gamma_2 \,e^{-2 \omega_2(R)}}{R^2 \mathfrak{g}^2(R,t)}
   +\frac{\lambda^2 \sin^2\gamma_2}{\mathfrak{g}^2(R,t)} \,,\\
   I_7 & =  \frac{r^4 \cos^4\gamma_2 \,e^{-4 \omega_2(R)}}{R^4 \mathfrak{g}^4(R,t)}
   +\frac{\lambda^4 \sin^2\gamma_2}{\mathfrak{g}^4(R,t)}   \,,\\
   I_8 & = \cos(\gamma_1-\gamma_2) 
   \left[\frac{r^2 \cos\gamma_1 \,\cos\gamma_2 \,e^{-2 \omega_2(R)}}{R^2 \,\mathfrak{g}^2(R,t)}
   +\frac{\lambda^2 \,\sin\gamma_1 \,\sin\gamma_2}{\mathfrak{g}^2(R,t)} \right] \,,\\
   I_9 & =  \cos^2(\gamma_1-\gamma_2)  \,.
\end{aligned}
\end{equation}
For an incompressible monoclinic solid, the Cauchy stress is given in \eqref{CauchyStress-Monoclinic}
where 
\begin{equation}
	W_i=\frac{\partial W}{\partial I_i}=\frac{\partial \We}{\partial I_i}+\frac{\partial \Wg}{\partial I_i}\,,
	\qquad i=1,2,4,5,\hdots 9\,.
\end{equation}
The non-zero Cauchy stress components are
\begin{equation}\label{Stress}
\begin{aligned}
	\sigma^{rr}(R,t) &= -\,p(R,t)
	+\frac{2 R^2 e^{2\omega_2(R)}\,\mathfrak{g}^4(R,t)}{\lambda^2\,r^2(R,t)} \,W_1
	+\left[\frac{2\,\mathfrak{g}^2(R,t)}{r^2(R,t)\,\lambda^2}\Big(r^2(R,t)+e^{2\omega_2(R)}R^2\lambda^2\Big)\right] W_2\,,
	\\[6pt]
	\sigma^{\theta\theta}(R,t) &= -\,\frac{p(R,t)}{r^2(R,t)}
	+\frac{e^{-2\omega_2(R)}}{R^2\,\mathfrak{g}^2(R,t)} \,W_1
	+\left(\frac{1}{R^2\,\mathfrak{g}^4(R,t)}\Big[e^{-2\omega_2(R)}\lambda^2+\frac{R^2\,\mathfrak{g}^6(R,t)}{r^2(R,t)\,	
	\lambda^2}\Big]\right) W_2\\
	&\quad
	+\frac{e^{-2\omega_2(R)}\cos^2\gamma_1}{R^2\,\mathfrak{g}^2(R,t)} \,W_4
	+\frac{2 e^{-4\omega_2(R)}\,r^2(R,t)\cos^2\gamma_1}{R^4\,\mathfrak{g}^4(R,t)} \,W_5
	\\
	&\quad
	+\frac{e^{-2\omega_2(R)}\cos^2\gamma_2}{R^2\,\mathfrak{g}^2(R,t)} \,W_6
	+\frac{2 e^{-4\omega_2(R)}\,r^2(R,t)\cos^2\gamma_2}{R^4\,\mathfrak{g}^4(R,t)} \,W_7
	\\
	&\quad
	+\frac{2 e^{-2\omega_2(R)}\cos\gamma_1\cos(\gamma_1-\gamma_2)\cos\gamma_2}{R^2\,\mathfrak{g}^2(R,t)} \,W_8\,,
	\\[6pt]
	\sigma^{\theta z}(R,t) &=
	\frac{\lambda e^{-\omega_2(R)}\,\sin(2\gamma_1)}{R\,\mathfrak{g}^2(R,t)} \,W_4
	+\frac{\lambda e^{-3\omega_2(R)}\,\sin(2\gamma_1)\left[r^2(R,t)+e^{2\omega_2(R)}R^2\lambda^2\right]}
	{R^3\,\mathfrak{g}^4(R,t)}	\, W_5
	\\
	&\quad
	+\frac{\lambda e^{-\omega_2(R)}\,\sin(2\gamma_2)}{R\,\mathfrak{g}^2(R,t)} \,W_6
	+\frac{\lambda e^{-3\omega_2(R)}\,\left[r^2(R,t)+e^{2\omega_2(R)}R^2\lambda^2\right]	
	\sin(2\gamma_2)}{R^3\,\mathfrak{g}^4(R,t)} \,W_7\\
	&\quad
	+\frac{\lambda e^{-\omega_2(R)}\,\left[\sin(2\gamma_1)+\sin(2\gamma_2)\right]}{R\,\mathfrak{g}^2(R,t)} \,W_8\,,\\[6pt]
	\sigma^{zz}(R,t) &= -\,p(R,t)
	+\frac{\lambda^2}{\mathfrak{g}^2(R,t)} \,W_1
	+\left(\frac{e^{-2\omega_2(R)}\,\lambda^2 r^2(R,t)}{R^2\,\mathfrak{g}^4(R,t)}
	+\frac{e^{2\omega_2(R)}\,R^2\,\mathfrak{g}^2(R,t)}{r^2(R,t)}\right) W_2
	\\
	&\quad
	+\frac{\lambda^2 \sin^2\gamma_1}{\mathfrak{g}^2(R,t)} \,W_4
	+\frac{2\lambda^4 \sin^2\gamma_1}{\mathfrak{g}^4(R,t)} \,W_5
	+\frac{\lambda^2 \sin^2\gamma_2}{\mathfrak{g}^2(R,t)} \,W_6
	+\frac{2\lambda^4 \sin^2\gamma_2}{\mathfrak{g}^4(R,t)} \,W_7\\
	&\quad
	+\frac{2\lambda^2 \sin\gamma_1 \sin\gamma_2 \cos(\gamma_1-\gamma_2)}{\mathfrak{g}^2(R,t)} \,W_8\,.
\end{aligned}
\end{equation}
The only nontrivial equilibrium equation $\sigma^{rr}{}_{,r}+\dfrac{1}{r}\sigma^{rr}-r\sigma^{\theta\theta}=0$ is simplified to read 
\begin{equation}
\frac{\partial}{\partial R}\sigma^{rr}(R,t) 
=  f(R,t) \,,
\end{equation}
where
\begin{equation}
\begin{aligned}
f(R,t)
&= \left( 
\frac{2 \lambda^2 e^{\omega_1(R)-\omega_2(R)} \,\mathfrak{g}(R,t)}{\lambda^3 R}
-\frac{2 R^3 e^{\omega_1(R)+3\omega_2(R)} \,\mathfrak{g}^7(R,t)}{\lambda^3 r^4(R,t)}
\right) W_1 \\[4pt]
&\quad +\left( 
\frac{2 \lambda e^{\omega_1(R)-\omega_2(R)}}{R \,\mathfrak{g}(R,t)}
-\frac{2 R^3 e^{\omega_1(R)+3\omega_2(R)} \,\mathfrak{g}^5(R,t)}{\lambda\, r^4(R,t)}
\right) W_2 \\[4pt]
&\quad +\frac{2 \cos^2\gamma_1 \,\mathfrak{g}(R,t)\, e^{\omega_1(R)-\omega_2(R)}}{\lambda R}\,W_4
+\frac{4 r^2(R,t)\cos^2\gamma_1 \, e^{\omega_1(R)-3\omega_2(R)}}{\lambda R^3 \,\mathfrak{g}(R,t)}\,W_5 \\[4pt]
&\quad +\frac{2 \cos^2\gamma_2 \,\mathfrak{g}(R,t)\, e^{\omega_1(R)-\omega_2(R)}}{\lambda R}\,W_6
+\frac{4 r^2(R,t)\cos^2\gamma_2 \, e^{\omega_1(R)-3\omega_2(R)}}{\lambda R^3 \,\mathfrak{g}(R,t)}\,W_7 \\[4pt]
&\quad +\frac{4 \cos\gamma_1 \cos\gamma_2 \cos(\gamma_1-\gamma_2)\,\mathfrak{g}(R,t)\, e^{\omega_1(R)-\omega_2(R)}}{\lambda R}\,W_8  \,.
\end{aligned}
\end{equation}
Integrating the above ODE from $R_i$ to $R$, recalling that $\sigma^{rr}(R_i,t)=-p_i(t)$, and using \eqref{Stress}$_1$, one obtains
\begin{equation}
\begin{aligned}
	-p(R,t)
	= -p_i(t)
	-\frac{2 R^2 e^{2\omega_2(R)}\,\mathfrak{g}^4(R,t)}{\lambda^2\,r^2(R,t)} \,W_1
	-\left[\frac{2\,\mathfrak{g}^2(R,t)}{r^2(R,t)\,\lambda^2}\Big(r^2(R,t)+e^{2\omega_2(R)}R^2\lambda^2\Big)\right] W_2 + \int_{R_i}^{R} f(\xi,t) d\xi \,.
\end{aligned}
\end{equation}
Therefore, the physical circumferential stress $\hat{\sigma}^{\theta\theta}=r^2\sigma^{\theta\theta}$, 
axial stress $\hat{\sigma}^{zz}=\sigma^{zz}$, and shear stress $\hat{\sigma}^{\theta z}=r \sigma^{\theta z}$ can be obtained. The boundary condition $\sigma^{rr}(R_o,t)=0$ implies that
\begin{equation}
\begin{aligned}
	\int_{R_i}^{R} f(\xi,t) d\xi = -p_i(t) \,.
\end{aligned} \label{eq:equilibrium-p}
\end{equation}

\begin{remark}
In modeling arterial growth, we consider a thick cylindrical shell that, in the absence of external forces, carries a distribution of eigenstrains. The artery is quasistatically loaded over the time interval $t\in[0,t_i]$ without bulk growth. Bulk growth starts at $t=t_i$.
For $t<t_i$, there is no growth and we have
\begin{equation} \label{initial-p}
\begin{aligned} 
	-\mathring{p}(R,t)
	& =-p_i(t) -\frac{2 R^2 e^{2 \omega_2(R)}}{\lambda^2 r^2(R,t)}\,W_1
	+\frac{2 \lambda^2 r^2(R,t)\, e^{-2 \omega_2(R)}}{R^2 }\,W_2   \\
	& \quad - \int_{R_i}^{R}  \frac{2 e^{\omega_1(\xi)-3 \omega_2(\xi)}  
	\left(\lambda^2 \,r^4(\xi,t) \,\xi^2 \,e^{2 \omega_2(\xi)} -\xi^6 \,e^{6\omega_2(\xi)} \right)}{\lambda^3 \,r^4(\xi,t) \,\xi^3 }
	\,W_1\,d\xi \\
	&\quad - \int_{R_i}^{R}  \frac{2 e^{\omega_1(\xi)-3 \omega_2(\xi)} 
	\left(\lambda^4 \,r^4(\xi,t)\, \xi^2\, e^{2\omega_2(\xi)} -\lambda^2 \,\xi^6 \,e^{6\omega_2(\xi)}  \right)}{\lambda^3 \,r^4(\xi,t)
	\, \xi^3 }\,W_2\, d\xi  \\
	& \quad -\int_{R_i}^{R}  \frac{2 \cos^2\gamma_1 \,e^{\omega_1(\xi)-\omega_2(\xi)}}{\lambda \,\xi }\,W_4\, d\xi 
	-\int_{R_i}^{R} \frac{4 r^2(\xi,t)\, \cos^2\gamma_1  \,e^{\omega_1(\xi)-3 \omega_2(\xi)}}{\lambda \, \xi^3}\,W_5\, d\xi  \\
	& \quad -\int_{R_i}^{R}  \frac{2 \cos^2\gamma_2  \,e^{\omega_1(\xi)-\omega_2(\xi)}}{\lambda  \,\xi}\,W_6\, d\xi 
	-\int_{R_i}^{R}  \frac{4 r^2(\xi,t)\, \cos^2\gamma_2 \,e^{\omega_1(\xi)-3\omega_2(\xi)}}{\lambda\, \xi^3 }\,W_7\, d\xi 	 \\
	& \quad -\int_{R_i}^{R}  \frac{4 \cos\gamma_1\, \cos\gamma_2\, \cos(\gamma_1-\gamma_2) 
	\,e^{\omega_1(\xi)-\omega_2(\xi)}}{\lambda \, \xi }\,W_8\, d\xi 
	\,.
\end{aligned}
\end{equation}
\end{remark}

\subsection{Growth equations} 

We write the total energy as $W=\We+\Wg$, where $\We=\We(X,I_1,I_2,I_4,\hdots,I_9)$ is the elastic energy and $\Wg$ is a growth energy that has the form given in \eqref{Anisotropic-Growth-Energy-Gradient}.
Note that, in general, $W_1, W_2, W_4, \hdots, W_9$ contain contributions from both the elastic and growth energies. For the growth energy specified in \eqref{Anisotropic-Growth-Energy-Gradient}, we have
\begin{equation}
\begin{aligned}
	\Wg_j:=\frac{\partial \Wg}{\partial I_j}
	=\frac{1}{2}(\mathfrak{g}-1)^2 \,\frac{\partial \kappa_{g}}{\partial I_j} 
	+\frac{1}{2}\frac{\partial \hat{\kappa}_{g}}{\partial I_j}\,\|\nabla\mathfrak{g}\|^2
	\,,\qquad j=1,2,4,\hdots,9\,.
\end{aligned}
\end{equation}
We assume the dissipation potential given in \eqref{Anisotropic-Dissipation} with its corresponding growth generalized force given in \eqref{Anisotropic-Growth-Forces}.
The variational principle \eqref{LD-Principle} tells us that 
\begin{equation} \label{Isotropic-Variational-P1}
	-\delta \We-\delta\Wg+p\,\delta J +B_{g}\,\delta\mathfrak{g}=0	\,.
\end{equation}
Recall that $J=\sqrt{\frac{\det\mathbf{g}}{\det\mathbf{G}}}\,\det\mathbf{F}$, and hence $\delta J=-\frac{1}{2}J\,\delta(\det\mathbf{G})=-\frac{1}{2}J\,\mathbf{G}^{-1}:\delta\mathbf{G}$. From $\mathbf{G}=\Fg^\star\mathring{\mathbf{G}}\Fg$, one obtains $\delta\mathbf{G}=\delta\Fg^\star\mathring{\mathbf{G}}\Fg+\Fg^\star\mathring{\mathbf{G}}\delta\Fg$. Thus, $\mathbf{G}^{-1}:\delta\mathbf{G}=\Fg^{-1}\mathring{\mathbf{G}}^{-1}\Fg^{-\star}:\left(\delta\Fg^\star\mathring{\mathbf{G}}\Fg+\Fg^\star\mathring{\mathbf{G}}\delta\Fg\right)=2\Fg^{-1}\mathring{\mathbf{G}}^{-1}\Fg^{-\star}:\Fg^\star\mathring{\mathbf{G}}\delta\Fg=2\Fg^{-1}:\delta\Fg$. Therefore
\begin{equation}
	\delta J= -J\,\Fg^{-1}:\delta\Fg= -\Fg^{-1}:\delta\Fg
	= -3\,\mathfrak{g}^{-1}\, \delta \mathfrak{g}
	\,.
\end{equation}
Thus, \eqref{Isotropic-Variational-P1} is simplified to read
\begin{equation} \label{Anisotropic-Variational-P2}
	\left(K\,\dot{\mathfrak{g}}+3p\,\mathfrak{g}^{-1}\right) \delta\mathfrak{g}
	=-\delta \We-\delta\Wg
	+\hat{B}_{g}\,\delta\mathfrak{g}	\,.
\end{equation}

\begin{remark}
When a Lagrange multiplier $p$ is introduced to enforce $J=1$, the variations $\delta r$ and $\delta\mathfrak{g}$ are treated as independent. Consider $I_1=C_{AB}\,G^{AB}$. If only $\mathfrak{g}$ varies, then $\delta I_1=C_{AB}\,\delta G^{AB}$. Thus, even though $C_{AB}$ depends on $r_{,R}$ as in \eqref{r1-g-relationship}, the variations $\delta r$ and $\delta\mathfrak{g}$ remain independent. In particular, this implies that for calculating variations of the invariants, \eqref{I1-I9-Definitions} cannot be used unless the dependence of $r$ on $\mathfrak{g}$ is taken into account.
\end{remark}

The variation of the invariants \eqref{I1-I9-Definitions} are written as
\begin{equation}
\begin{aligned}
\delta I_1 &= 
\left[
-\frac{2 r^2(R,t) e^{-2\omega_2(R)}}{R^2 \mathfrak{g}^3(R,t)}
-\frac{2\lambda^2}{\mathfrak{g}^3(R,t)}
-\frac{2 R^2 e^{2\omega_2(R)}\,\mathfrak{g}^3(R,t)}{r^2(R,t) \lambda^2}
\right]\delta\mathfrak{g}(R,t) = \mathfrak{I}_1 \delta\mathfrak{g}(R,t)\,, \\
\delta I_2 &= 
\left[
-\frac{4 \lambda^2 r^2(R,t) e^{-2\omega_2(R)}}{R^2 \mathfrak{g}^5(R,t)}
-\frac{4\left(r^2(R,t) + \lambda^2 R^2 e^{2\omega_2(R)}\right)\mathfrak{g}(R,t)}{r^2(R,t) \lambda^2}
\right]\delta\mathfrak{g}(R,t)= \mathfrak{I}_2 \delta\mathfrak{g}(R,t)\,, \\
\delta I_4 &= 
-\frac{2\left(r^2(R,t) e^{-2\omega_2(R)}\cos^2\gamma_1 
+ \lambda^2 R^2 \sin^2\gamma_1\right)}{R^2 \mathfrak{g}^3(R,t)}
\,\delta\mathfrak{g}(R,t)= \mathfrak{I}_4 \delta\mathfrak{g}(R,t)\,, \\
\delta I_5 &= 
-\frac{4\left(r^4(R,t) e^{-4\omega_2(R)}\cos^2\gamma_1 
+ \lambda^4 R^4 \sin^2\gamma_1\right)}{R^4 \mathfrak{g}^5(R,t)}
\,\delta\mathfrak{g}(R,t)= \mathfrak{I}_5 \delta\mathfrak{g}(R,t)\,, \\
\delta I_6 &= 
-\frac{2\left(r^2(R,t) e^{-2\omega_2(R)}\cos^2\gamma_2 
+ \lambda^2 R^2 \sin^2\gamma_2\right)}{R^2 \mathfrak{g}^3(R,t)}
\,\delta\mathfrak{g}(R,t)= \mathfrak{I}_6 \delta\mathfrak{g}(R,t)\,, \\
\delta I_7 &= 
-\frac{4\left(r^4(R,t) e^{-4\omega_2(R)}\cos^2\gamma_2 
+ \lambda^4 R^4 \sin^2\gamma_2\right)}{R^4 \mathfrak{g}^5(R,t)}
\,\delta\mathfrak{g}(R,t)= \mathfrak{I}_7 \delta\mathfrak{g}(R,t)\,, \\
\delta I_8 &= 
\left[
-\frac{2 e^{-2\omega_2(R)} \cos(\gamma_1-\gamma_2)\,
\left(r^2(R,t) \cos\gamma_1\cos\gamma_2 
+ \lambda^2 R^2 e^{2\omega_2(R)} \sin\gamma_1 \sin\gamma_2\right)}
{R^2 \mathfrak{g}^3(R,t)}
\right]\delta\mathfrak{g}(R,t)= \mathfrak{I}_8 \delta\mathfrak{g}(R,t)\,.
\end{aligned}
\end{equation}
Note that
\begin{equation}
\begin{aligned}
	\delta\Wg &=
	\frac{\partial\Wg}{\partial \mathfrak{g}}\delta \mathfrak{g}
	+\frac{\partial\Wg}{\partial \nabla\mathfrak{g}}\cdot \nabla\delta \mathfrak{g} 
	+\Wg_1 \,\delta I_1+\Wg_2 \,\delta I_2+\Wg_4 \,\delta I_4+\hdots+\Wg_2 \,\delta I_9 \\
	& =	 \kappa_{g}\,(\mathfrak{g}-1)\delta \mathfrak{g}
	+\hat{\kappa}_{g}\,\nabla\mathfrak{g}\cdot \delta \nabla\mathfrak{g}
	+\Wg_1 \,\delta I_1+\Wg_2 \,\delta I_2+\Wg_4 \,\delta I_4+\hdots+\Wg_9 \,\delta I_9
	\,,
\end{aligned}
\end{equation}
where 
\begin{equation}
	\Wg_i:=\frac{\partial \Wg}{\partial I_i}
	=\frac{1}{2}(\mathfrak{g}-1)^2 \frac{\partial \kappa_g}{\partial I_i} 
	+\frac{1}{2}\frac{\hat{\kappa}_g}{\partial I_i}\,\|\nabla\mathfrak{g}\|^2
	\,,\qquad i=1,2,4,\hdots,9 \,.
\end{equation}
Recall that for an incompressible monoclinic solid $\We=\We(X,I_1,I_2,I_4,\hdots,I_9)$. Thus
\begin{equation}
	\delta \We= \We_1 \,\delta I_1+\We_2 \,\delta I_2+\We_4 \,\delta I_4+\hdots+\We_9 \,\delta I_9
	\,.
\end{equation}
Therefore
\begin{equation}
\begin{aligned}
	\delta\We+\delta\Wg & =	\kappa_{g}\,(\mathfrak{g}-1)\delta \mathfrak{g}
	+\hat{\kappa}_{g}\,\nabla\mathfrak{g}\cdot \delta \nabla\mathfrak{g} 
	\\
	& \quad +(\We_1+\Wg_1) \,\delta I_1+(\We_2+\Wg_2) \,\delta I_2+(\We_4+\Wg_4) \,\delta I_4
	+\hdots+(\We_9+\Wg_9) \,\delta I_9\,.
\end{aligned}
\end{equation}
Eq.~\eqref{Anisotropic-Variational-P2} is simplified to read
Simplifying \eqref{Anisotropic-Variational-P2} and using the variational principle \eqref{LD-Principle} gives us the following growth equation
\begin{equation} \label{GrowthEquation-General}
\begin{aligned}
K\,\dot{\mathfrak{g}}
+3p\,\mathfrak{g}^{-1}
&=\hat{\kappa}_{g}\,\Delta\mathfrak{g}
-\kappa_{g}(\mathfrak{g}-1) +\hat{B}_{g} \\[4pt]
&\quad +(\We_1+\Wg_1) \mathfrak{I}_1 
+(\We_2+\Wg_2) \mathfrak{I}_2 
 +(\We_4+\Wg_4) \mathfrak{I}_4 
 +(\We_5+\Wg_5) \mathfrak{I}_5 
 +(\We_6+\Wg_6) \mathfrak{I}_6 \\[4pt]
&\quad +(\We_7+\Wg_7) \mathfrak{I}_7 
+(\We_8+\Wg_8) \mathfrak{I}_8\,,
\end{aligned}
\end{equation}
along with the boundary condition $\mathfrak{g}'(R_0,t)=0$.

We would like to make sure that $\mathfrak{g}(R,t)=1$ for $t<t_i$. It should be noted that a configurational force is required to control the growth equation at the onset of the growth process. More precisely, we start from an undeformed artery and quasi-statically load it until the normal physiological lumen pressure is reached, at which point the growth process begins. The growth equation above can then be obtained from \eqref{GrowthEquation-General} by the following choice of the configurational force $\hat{B}_{g}$:
\begin{equation}
\begin{aligned}
\hat{B}_{g} &= 3\,\mathring{p}\,\mathfrak{g}^{-1} \\
&\quad -(\We_1+\Wg_1) \mathring{\mathfrak{I}}_1 
-(\We_2+\Wg_2) \mathring{\mathfrak{I}}_2 
 -(\We_4+\Wg_4) \mathring{\mathfrak{I}}_4 
 -(\We_5+\Wg_5) \mathring{\mathfrak{I}}_5 
 -(\We_6+\Wg_6) \mathring{\mathfrak{I}}_6 \\[4pt]
&\quad -(\We_7+\Wg_7) \mathring{\mathfrak{I}}_7 
-(\We_8+\Wg_8) \mathring{\mathfrak{I}}_8
\,,
\end{aligned}
\end{equation}
where $\mathring{\mathfrak{I}}_i $ ($i$=1,2,4,5,6,7,8) stand for the functions ${\mathfrak{I}}_i$ for $g=1$.
Then the growth equations take the final form
\begin{equation}
\begin{aligned}
& K\,\dot{\mathfrak{g}}
+3(p-\mathring{p})\,\mathfrak{g}^{-1}
=\hat{\kappa}_{g}\,\Delta\mathfrak{g}
-\kappa_{g}(\mathfrak{g}-1) \\[4pt]
&\quad +(\We_1+\Wg_1) (\mathfrak{I}_1 - \mathring{\mathfrak{I}}_1 ) 
+(\We_2+\Wg_2) (\mathfrak{I}_2 - \mathring{\mathfrak{I}}_2 )  
 +(\We_4+\Wg_4) (\mathfrak{I}_4 - \mathring{\mathfrak{I}}_4)  
 +(\We_5+\Wg_5) (\mathfrak{I}_5 - \mathring{\mathfrak{I}}_5 )  
  \\[4pt]
&\quad +(\We_6+\Wg_6) (\mathfrak{I}_6 - \mathring{\mathfrak{I}}_6 )  +(\We_7+\Wg_7) (\mathfrak{I}_7 - \mathring{\mathfrak{I}}_7 )  
+(\We_8+\Wg_8) (\mathfrak{I}_8 - \mathring{\mathfrak{I}}_8 ) \,,
\end{aligned} \label{eq:growth-eq-final}
\end{equation}
where $\mathring{p}$ is given in \eqref{initial-p}.
It can be observed from the numerical results in \S\ref{Sec:NumericalResults}  that for $t<t_i$, $\mathfrak{g}(R,t)=1$.

\subsection{The bulk growth initial-boundary value problem}

For the artery, the force $F(t)$ at the two ends of the bar ($Z=0, L$) and the internal pressure are known. Thus the unknown fields are $r_i(t)$, $\lambda(t)$ and $\mathfrak{g}(R,t)$. They are obtained from solving the  initial-value problem that consists of the growth equation (\ref{eq:growth-eq-final}), equilibrium equation (\ref{eq:equilibrium-p}), axial equilibrium equation  
\begin{equation}
    F(t)=2\pi \int_{r_i(t)}^{r_o(t)}\sigma^{zz}(r,t)r\,dr = \frac{2\pi}{\lambda(t)} \int_{R_i}^{R_o} \sigma^{zz}(R,t)\,, 
\label{axial-equilibrium}
\end{equation}
along with initial conditions $\mathfrak{g}(R,0)=1$ and $r_i(0)=R_i$.


\begin{remark}
The local balance of mass \eqref{Mass-Balance-Local} is simplified to read
\begin{equation} 
	\dot{\rho}_0(R,t) + 3\mathfrak{g}^{-1}(R,t)\dot{\mathfrak{g}}(R,t)\,\rho_0(R,t) = S_m(R,t) \,.
\end{equation}
This ODE determines the material mass density $\rho_0(R,t)$.
\end{remark}

\section{Numerical Results}
\label{Sec:NumericalResults}

The system of nonlinear integro-differential equations derived in the preceding section is solved numerically by employing the finite difference method. In the present analysis, we consider a rabbit carotid artery, whose elastic response has been experimentally characterized in previous studies. The experimental data are well represented by the Fung-type, specifically the Holzapfel–Gasser–Ogden (HGO) \citep{holzapfel2004comparison, gasser2006hyperelastic}, hyperelastic constitutive model, for which
\begin{equation} \label{Fung-material}
	\We=\frac{\mu_1}{2} (I_1-3)
	+\frac{k_1}{2 k_2} \left( e^{\,{k_2} (I_4 -1)^2} -1 \right)
    + \frac{k_1}{2 k_2} \left( e^{\,{k_2} (I_6 -1)^2} -1 \right)
	\,,
\end{equation}	
where $\mu_1$, $k_1$, and $k_2$ are positive material constants. The material constants are specified as follows. For the inner medial layer of the artery, they take the values $\mu_1 = 3~\text{kPa}$ and $k_1 = 2.36~\text{kPa}$. For the outer adventitial layer, the corresponding values are $\mu_1 = 0.3~\text{kPa}$ and $k_1 = 0.56~\text{kPa}$ \citep{holzapfel2004data}. We will vary the values of constant $k_2$ for the two layers. The geometric dimensions of the arterial wall are defined by an inner radius of the media $R_i = 0.71~\text{mm}$, an outer radius of the media $R_m = 0.97~\text{mm}$, and an outer radius of the adventitia $R_o = 1.10~\text{mm}$.
The collagen fiber orientations in the medial layer are taken as $\gamma_1 = 29^{\circ}$ and $\gamma_2 = -29^{\circ}$, whereas in the adventitial layer they are $\gamma_1 = 62^{\circ}$ and $\gamma_2 = -62^{\circ}$, but will also be varied. To examine the role of anisotropy, results for isotropic elasticity are also considered by setting $k_1 = 0$. The normotensive internal pressure is prescribed as $p_0 = 1 \times (\mu_1 + k_1)$, while the hypertensive pressure increment is given by $\Delta p_0 = 0.25 \times (\mu_1 + k_1)$. In the isotropic elasticity case, pressures of $p_0 = 0.4 \times \mu_1$ and $\Delta p_0 = 0.1 \times \mu_1$ are adopted such that the inner radius under normotensive and hypertensive conditions closely matches that obtained for the anisotropic case. The parameter $K$, which appears in the dissipation potential and governs the characteristic time scale associated with the growth process, is assigned a value of $K = 0.03$ .

\begin{figure}[h!]
	\begin{center}
	\vskip 0.3in
		\includegraphics[width=0.7\textwidth]{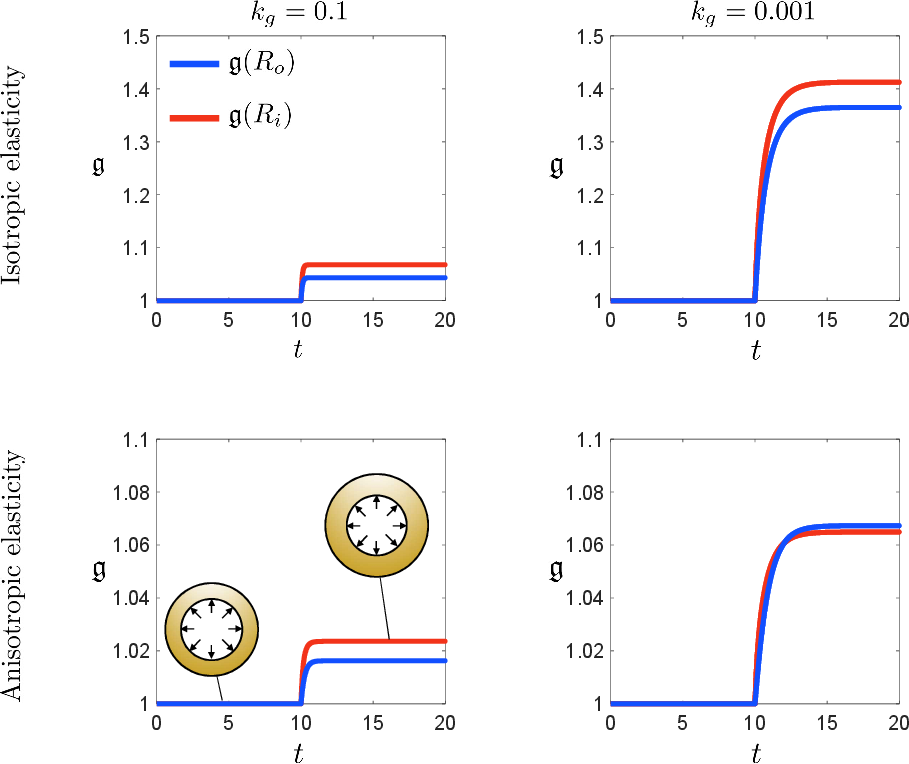}
	\end{center}
	\vspace{-0.2cm} \caption{\textit{Isotropic growth evolution in a single layer artery. The growth parameter $\mathfrak{g}$ is plotted at the inner ($R = R_i$) and outer ($R = R_o$) radii as a function of time $t$. Results are shown for two values of the growth energy parameter, $k_g = 0.1$ and $0.001$, and for both isotropic and anisotropic elastic responses. }}
	\label{Fig1}
\end{figure}

\subsection{Growth in a single-layer artery}

Since the medial layer constitutes the primary load-bearing component of the arterial wall and exhibits the most pronounced growth response, we first analyze a simplified single-layer model of the artery, neglecting the adventitial layer \citep{Rodriguez2007}. This assumption also facilitates a clearer illustration of the fundamental behavior of the proposed model.
Figure~\ref{Fig1} presents the temporal evolution of the isotropic growth parameter $\mathfrak{g}$ as a function of the nondimensional time variable $t$ for two representative values of the growth energy parameter, $k_g = 0.1$ and $k_g = 0.001$.
In addition to the results obtained for the anisotropic elastic response of the arterial wall, those corresponding to an isotropic elasticity assumption ($k_1 = 0$) are also shown for comparison. The results reveal that the magnitude of growth is governed by the competition between the growth energy and the strain energy stored in the tissue. In contrast to conventional models of volumetric growth \citep{Lubarda2002}, no explicit upper bound on the growth variable is prescribed. Instead, growth naturally approaches an asymptotic steady-state value as a result of this intrinsic energetic competition. Notably, the asymptotic growth value is found to be independent of the dissipation parameter $K$.

Variations in the growth energy constant $k_g$ significantly influence the asymptotic magnitude of growth. For smaller values of $k_g$, the energetic cost associated with growth decreases, thereby allowing for greater growth. However, the extent of growth eventually saturates for sufficiently small $k_g$, indicating that beyond a threshold (approximately $k_g < 0.001$ for the present arterial configuration), further reductions in $k_g$ do not lead to appreciable changes in the maximum attainable growth. Thus, despite uncertainties in the precise calibration of $k_g$ and, more broadly, the growth energy function, the proposed formulation robustly predicts a physically realistic, bounded growth response.

An additional noteworthy observation concerns the influence of anisotropic elasticity. The inclusion of anisotropy results in a more constrained isotropic growth response, with the growth parameter $\mathfrak{g}$ saturating at approximately 10\% for low values of $k_g$. Although this behavior may differ when anisotropic growth mechanisms are incorporated, these results underscore the important role of three-dimensional structural anisotropy in regulating bulk growth. This aspect is further explored through a parametric study presented in Figure~\ref{Fig4}.

\begin{figure}[h!]
	\begin{center}
	\vskip 0.3in
		\includegraphics[width=0.7\textwidth]{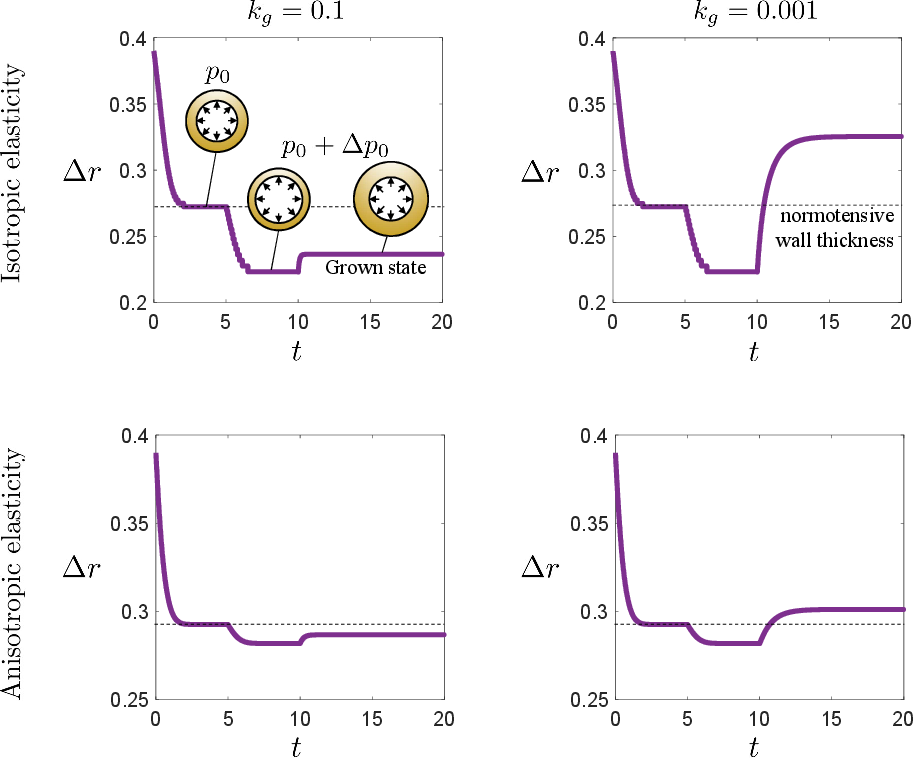}
	\end{center}
	\vspace{-0.2cm} \caption{Evolution of arterial wall thickness, $\Delta r$, under three conditions: normotensive (internal pressure $p_0$), hypertensive (internal pressure $p_0 + \Delta p_0$), and growth under hypertensive loading. Results are shown for isotropic and anisotropic elasticity, considering two values of the growth energy parameter, $k_g$. }
	\label{Fig2}
\end{figure}
Figure~\ref{Fig2} depicts the corresponding temporal evolution of the arterial wall thickness, denoted by $\Delta r$, as a function of the time variable $t$, for $k_g = 0.1$ and $k_g = 0.001$. The evolution process may be observed in three distinct stages. In the first stage, the cylindrical arterial wall is subjected to the normotensive pressure $p_0$ and the corresponding axial force $F$. The second stage corresponds to the onset of a hypertensive condition, during which the pressure increases to $p_0 + \Delta p_0$. The third stage captures the subsequent gradual evolution of wall thickness resulting from the growth process under sustained hypertensive loading.

For higher values of $k_g$, the growth-induced change in wall thickness during the third stage exhibits a modest increase relative to the hypertensive state, yet $\Delta r$ remains below its initial (normotensive) value. In contrast, for smaller values of $k_g$, the reduced energetic cost associated with growth permits $\Delta r$ to exceed the normotensive wall thickness. Under isotropic elasticity, where the model predicts substantially higher levels of growth, the wall thickening can reach magnitudes exceeding 20\% relative to the normotensive configuration. Conversely, in the case of anisotropic elasticity, the enhanced directional stiffness provides a regulatory effect, leading to a more moderate increase in wall thickness.
For comparison, experimental observations in both animal and human studies have reported hypertensive wall thickening in the range of approximately 10–50\% \citep{kuhl2014artery, fridez2002adaptation, wiener1977data, boutouyrie1999data, schofield2002data}.

\begin{figure}[h!]
	\begin{center}
	\vskip 0.3in
		\includegraphics[width=0.7\textwidth]{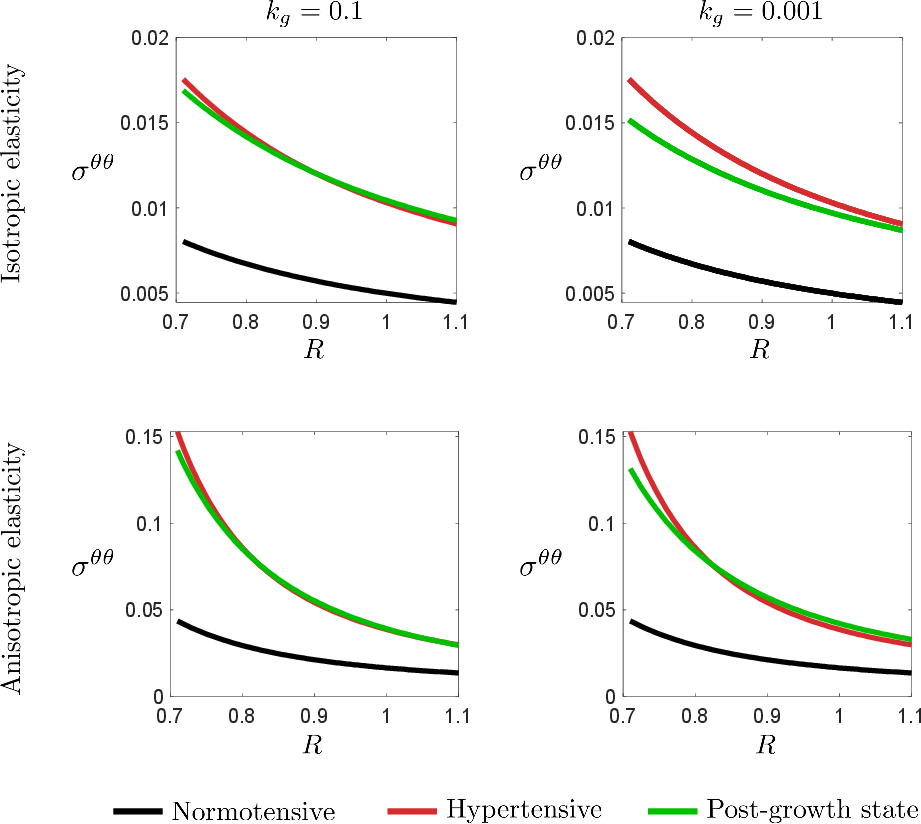}
	\end{center}
	\vspace{-0.2cm} \caption{Hoop (circumferential) stress, $\sigma^{\theta\theta}$, radial profile along the wall thickness plotted as a function of the radial coordinate $R$. Results are plotted at three states: normotensive state ($t = 5$), the hypertensive state ($t = 10$), and the post-growth state ($t = 20$). They are shown for isotropic and anisotropic elasticity, considering two values of the growth energy parameter, $k_g$.}
	\label{Fig3}
\end{figure}
Figure~\ref{Fig3} presents the radial profiles of the hoop (circumferential) stress, $\sigma^{\theta\theta}$, plotted as a function of the radial coordinate $R$, for the normotensive state ($t = 5$), the hypertensive state ($t = 10$), and the post-growth state ($t = 20$). Growth influences the hoop stress distribution, partially restoring it toward the normotensive condition and leading to a more uniform stress distribution across the arterial wall thickness, consistent with observations \citep{chaudhry1997residual}.
However, even substantial growth in the present isotropic growth analysis results in only modest reductions in the hoop stress, such that the overall stress state remains considerably elevated relative to the normotensive configuration. It should be noted that an anisotropic growth formulation could produce a more pronounced reduction in the circumferential stress \citep{Goriely2017}. Nevertheless, even in such cases, the hoop stress is not expected to return precisely to its normotensive magnitude at every point through the wall thickness.

\begin{figure}[h!]
	\begin{center}
	\vskip 0.3in
		\includegraphics[width=1\textwidth]{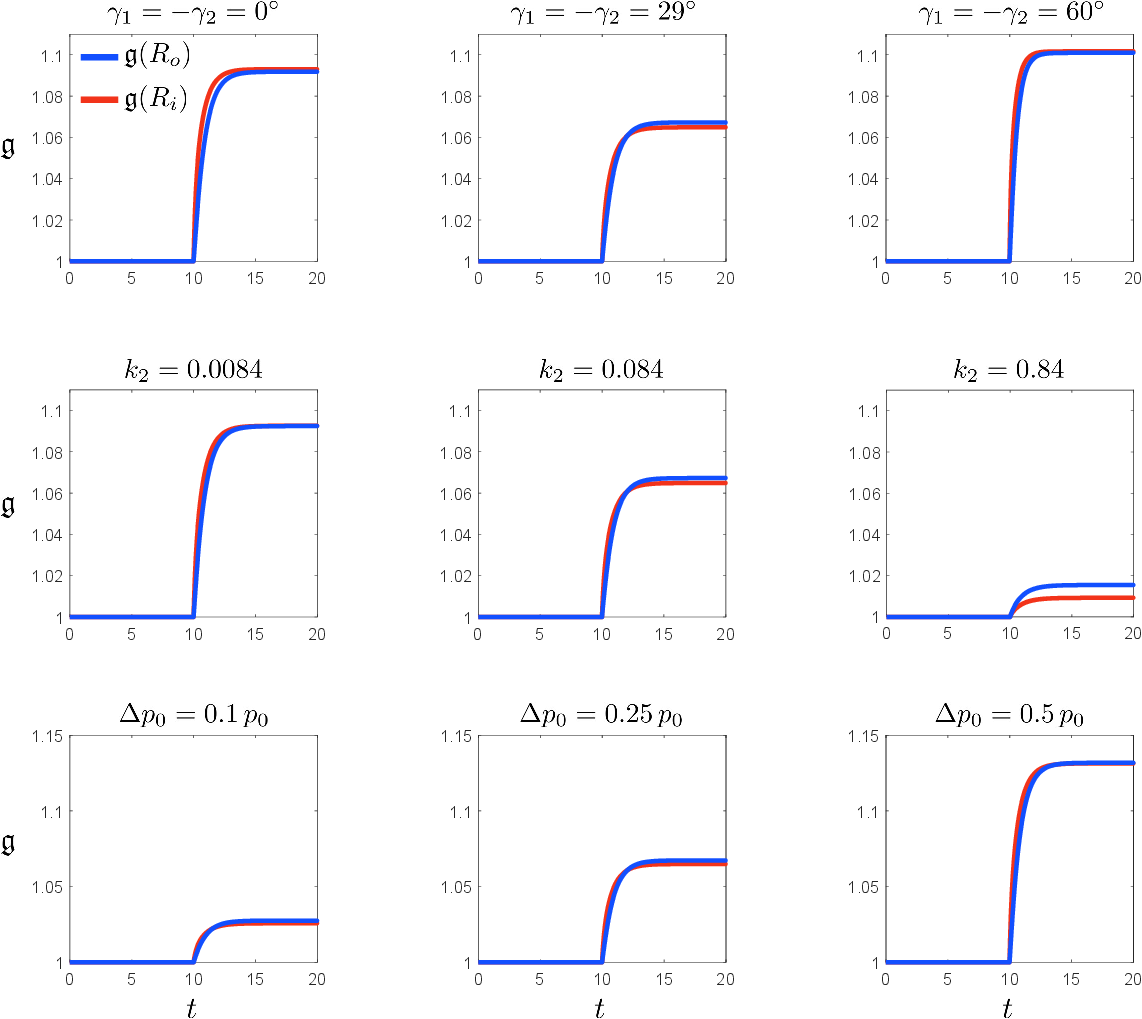}
	\end{center}
	\vspace{-0.2cm} \caption{Parametric study assessing the role of fiber orientation angles $(\gamma_1, \gamma_2)$, the material parameter $k_2$, and the pressure increment $\Delta p_0$ on growth parameter $\mathfrak{g}$.}
	\label{Fig4}
\end{figure}
To conclude this section, Figure~\ref{Fig4} presents a comprehensive parametric study in which the fiber orientation angles $(\gamma_1, \gamma_2)$, the material parameter $k_2$, and the pressure increment $\Delta p_0$ are systematically varied to assess their respective influences on arterial growth. The temporal evolution of the isotropic growth parameter $\mathfrak{g}$ is plotted with respect to time $t$ at both the inner and outer radii, $R = R_i$ and $R = R_o$, respectively. Three representative fiber orientation configurations, $\gamma_1 = -\gamma_2 = 0^{\circ}$, $29^{\circ}$, and $60^{\circ}$, are examined. The results indicate that the growth parameter exhibits a non-monotonic dependence on the fiber orientation angles under identical internal pressure and axial force conditions. Specifically, $\mathfrak{g}$ attains a lower steady-state value for $\gamma_1 = -\gamma_2 = 29^{\circ}$ compared to both the $0^{\circ}$ and $60^{\circ}$ cases. This finding may provide valuable insight into the well-documented phenomenon of collagen fiber remodeling in arterial walls \citep{Driessen2003, Hariton2007}. Remodeling and growth are intricately connected processes in arterial walls \citep{Baaijens2010}. Although a fully coupled analysis of growth and remodeling lies beyond the scope of the present work, related aspects of remodeling have been investigated previously within a similar variational framework \citep{KumarYavari2023}.

Next, the influence of the material parameter $k_2$ in the HGO hyperelastic model (\ref{Fung-material}) is analyzed by considering three representative values: $k_2 = 0.0084$, $0.084$, and $0.84$, under identical loading conditions. This parameter governs the degree of exponential stiffening in the nonlinear constitutive relation. As expected, larger values of $k_2$ result in stronger stiffening behavior, leading to a lower incentive to grow under force-controlled loading and, consequently, reduced growth.

Finally, the effect of the pressure increment is examined by considering three levels of hypertensive loading: $\Delta p_0 = 0.1p_0$, $0.25p_0$, and $0.5p_0$. As anticipated, higher pressure increments lead to more pronounced growth. Interestingly, for the anisotropic arterial material considered, an increase of 50\% in internal pressure produces only about a 15\% increase in the growth parameter $\mathfrak{g}$, corresponding to an approximately 15\% increase in wall thickness. This result highlights the moderating influence of anisotropy on the growth response under elevated mechanical loads.

\begin{figure}[h!]
	\begin{center}
	\vskip 0.3in
		\includegraphics[width=0.7\textwidth]{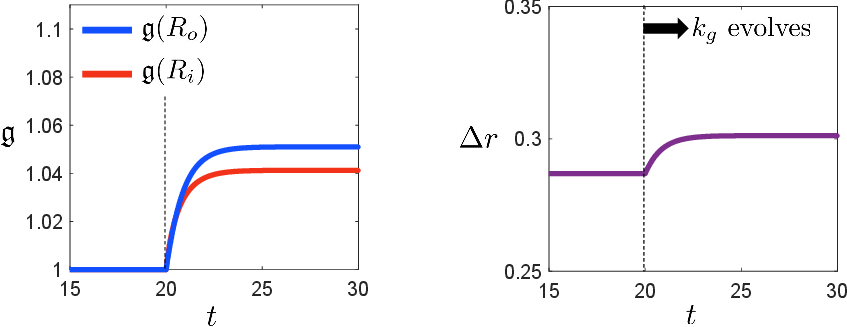}
	\end{center}
	\vspace{-0.2cm} \caption{Growth driven by the evolution in growth energy parameter $k_g$ from 0.1 to 0.001 at $t=20$.}
	\label{Fig5}
\end{figure}

\subsection{Growth generated by internal changes in the growth incentives}

In the preceding section, growth was initiated through an increase in the internal pressure within the arterial wall. A distinctive and novel aspect of the present variational formulation, however, is its ability to predict growth that may also be driven by the evolution of the growth energy itself, independent of changes in the applied loads. There exists some experimental and theoretical evidence supporting this type of intrinsic growth incentive in the literature \citep{humphrey2016central}. Specifically, it has been proposed that, under certain physiological conditions, hypertension may be preceded by an increase in structural stiffness, loosely defined as the product of wall thickness and material stiffness.

Within the context of the present framework, an evolution in growth energy (or equivalently, in the effective material elasticity) can lead to a change in arterial geometry, typically manifested as an increase in both the inner radius and the wall thickness. Such geometric adaptation may, in turn, alter the hemodynamic environment, resulting in elevated lumen pressure. This elevation could potentially establish a negative feedback mechanism in which elevated pressure drives further growth.

To illustrate this capability of the variational formulation, we consider a simplified numerical example in which the artery is subjected to constant internal pressure and axial force. At a given instant, the growth energy parameter $k_g$ is assumed to undergo an abrupt reduction from $0.1$ to $0.001$. This change alone, without any modification to the external loading, induces an evolution in the growth parameter $\mathfrak{g}$ and the corresponding wall thickness, thereby demonstrating growth purely driven by variations in the growth energy. A more comprehensive investigation of this phenomenon, encompassing different functional forms of the growth energy and potentially incorporating multiscale modeling, is needed to better understand this feature of the proposed variational framework.

\begin{figure}[h!]
	\begin{center}
	\vskip 0.3in
		\includegraphics[width=0.7\textwidth]{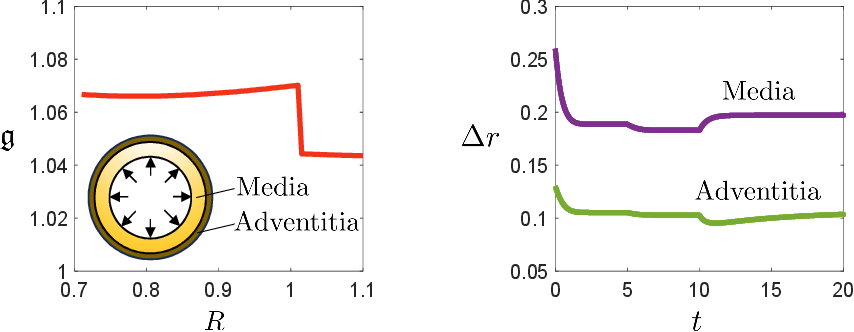}
	\end{center}
	\vspace{-0.2cm} \caption{Isotropic growth in a two-layered artery. Radial distribution of the growth, $\mathfrak{g}$ (left), and the evolution of wall thickness, $\Delta r$, in the media and adventitia (right).}
	\label{Fig6}
\end{figure}

\subsection{Effect of eigenstrains and differential growth between arterial layers}

Next, the artery is modeled as a two-layered cylindrical shell, with the medial layer occupying the region $0.71~\text{mm} \leq R < 0.97~\text{mm}$ and the adventitial layer extending from $0.97~\text{mm} \leq R \leq 1.10~\text{mm}$ \citep{holzapfel2004data}. The growth energy parameter $k_g$ is assumed to be identical in both layers. Figure~\ref{Fig6} presents the resulting distribution of the growth parameter $\mathfrak{g}$ at the conclusion of the growth process, alongside the corresponding temporal evolution of the wall thickness, $\Delta r$. The results indicate reduced growth in the comparatively softer adventitial layer. The thickness of the medial layer increases relative to its normotensive value, whereas the greater differential growth in the media induces a slight decrease in the adventitial thickness compared to its normotensive state. This more modest growth in the adventitia, coupled with the negligible change in wall thickness, is consistent with experimental observations.

\begin{figure}[h!]
	\begin{center}
	\vskip 0.3in
		\includegraphics[width=1\textwidth]{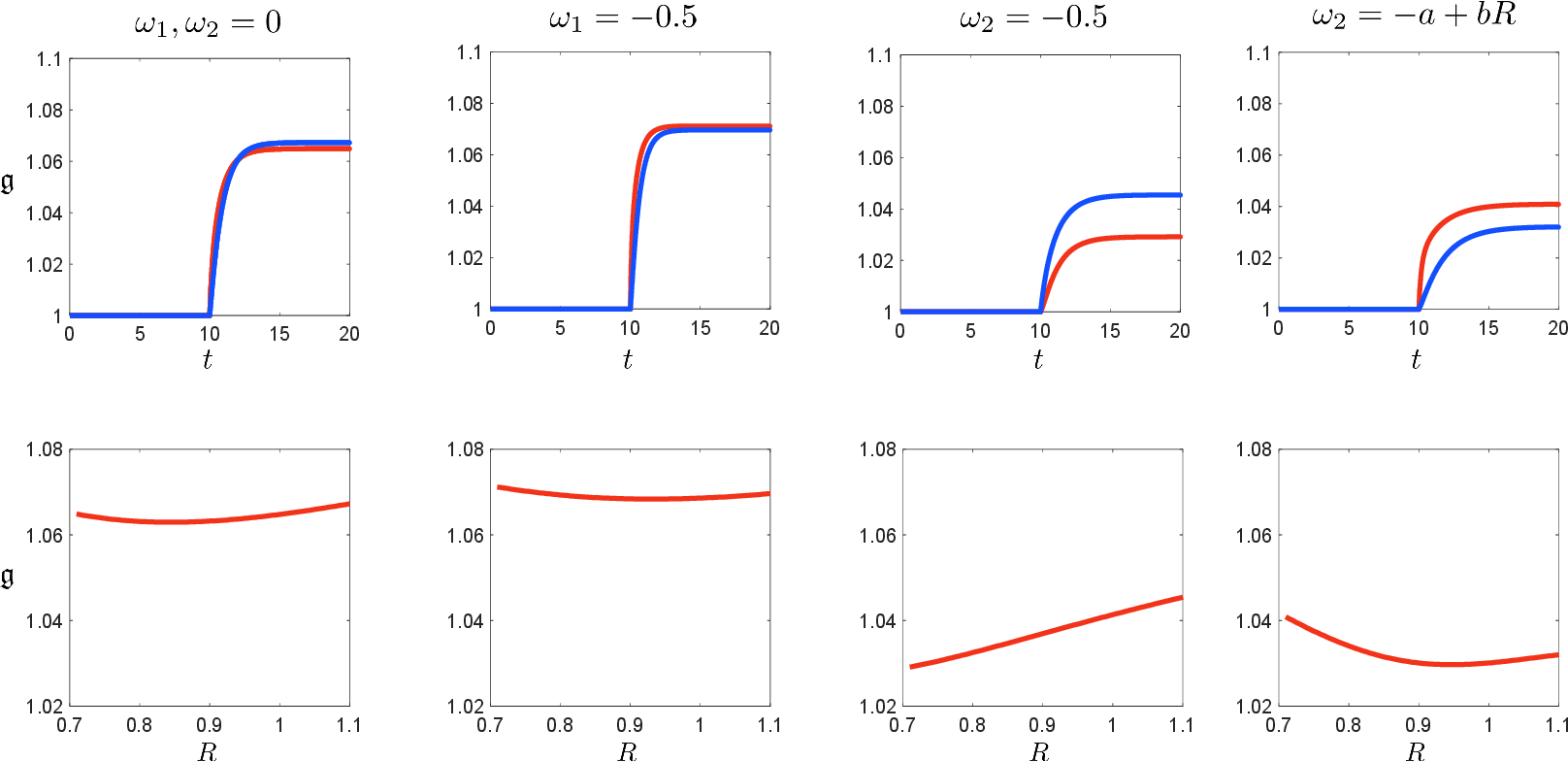}
	\end{center}
	\vspace{-0.2cm} \caption{Effect of radial eigenstrain, $\omega_1$ and circumferential eigenstrain, $\omega_2$ on the evolution of isotropic growth. The growth parameter is plotted as a function of time $t$ and as a function of radial coordinate $R$ (at $t=20$) in the top and bottom rows, respectively.  For the non-uniform eigenstrain distribution, the constants are $a=1.9$ and $b=1.54$.}
	\label{Fig7}
\end{figure}

Finally, Figure~\ref{Fig7} illustrates the evolution of the growth parameter when radial and circumferential eigenstrains, denoted $\omega_1$ (radial) and $\omega_2$ (circumferential), respectively, are incorporated. It is well established that arteries contain residual stresses/strains, which manifest as differences in the resting lengths of layers upon cutting \citep{vaishnav1983estimation, holzapfel2007, cardamone2009origin}. The eigenstrains in the present model are intended to account for the effects of prior developmental processes and physiological growth that generate these residual stresses. The results demonstrate that the introduction of circumferential eigenstrains increases the non-uniformity of growth across the wall thickness, while slightly reducing the overall magnitude of growth. In contrast, the inclusion of radial eigenstrains has little effect, yielding growth patterns comparable to those in the no-eigenstrain case. Finally, results for a non-uniform distribution of eigenstrains, $\omega_2= -a + b R$ with $a=1.9$ and $b=1.54$ are also shown. They produce qualitatively similar outcomes but affect the non-uniformity of growth across the thickness. Introducing the gradient effect through a non-zero value of the parameter $\hat{\kappa}_{g}$ also has a similar effect.

\section{Conclusions}  \label{Sec:Conclusions}

In this paper, we formulated a general theoretical framework for the mechanics of bulk growth based on the Lagrange--d'Alembert principle. The formulation is variational in nature and unifies elasticity, growth, and dissipation within a two-potential approach. The key idea is to introduce, in addition to the elastic energy and dissipation potential, a growth energy function that drives the evolution of the growth tensor through a configurational force analogous to the Eshelby stress. This framework allows one to describe growth as a dissipative process in the space of internal metrics, where the evolution law emerges naturally from the balance of elastic and configurational forces. The resulting governing equations provide a systematic and thermodynamically consistent description of bulk growth in anisotropic solids and the notion of mechanical homoestasis, and in particular, offer a basis for studying arterial growth under physiological and pathological conditions.

The numerical simulations were carried out for a rabbit carotid artery with particular emphasis on growth in the medial layer. The Fung-type (Holzapfel-Gasser-Ogden) hyperelastic model is utilized; however, the equations are provided for an arbitrary monoclinic solid.  We adopted a simple form of the growth energy, independent of deformation and stress, and quadratic in the growth factor. Even with this simple choice, the results demonstrate that the variational formulation produces a bounded and physically reasonable growth response. The extent of growth is modulated by multiple factors, including the growth-energy parameter, the magnitude and orientation of anisotropy, pressure elevation above the normotensive state, and any pre-existing eigenstrains. Consistent with physiological expectations, growth leads to an increase in arterial wall thickness and promotes a more uniform distribution of hoop stress across the wall. When growth in both the media and adventitia is considered, the results further reveal that differential growth between the two layers can induce a modest reduction in adventitial thickness. A particularly noteworthy observation from the results is that the variational formulation permits growth to arise solely from alterations in the tissue’s internal structure, manifested as changes in growth energy, even in the absence of any increase in lumen pressure. Such a cause for growth has been hypothesized in the literature.

In future work, incorporating anisotropic growth will be essential for capturing the inherently direction-dependent growth observed in arterial tissues. A related goal is to establish a better understanding of the form of the growth energy. In particular, understanding how the macroscopic growth energy relates to the energetics of underlying cellular processes will provide critical insight and ensure that the variational framework remains grounded in biophysical mechanisms.

\section*{Acknowledgement}

This work was partially supported by NSF -- Grant No. CMMI 1939901.

\bibliographystyle{abbrvnat}
\bibliography{ref}

\end{document}